\documentclass[twocolumn,aps,prd,preprintnumbers,showpacs,superscriptaddress,nofootinbib,amsmath,amssymb,floats,floatfix,showkeys,notitlepage,longbibliography]{revtex4-1}
\usepackage{subfigure} 
\usepackage{comment}
\usepackage{graphicx}
\usepackage{palatino}
\usepackage[commandnameprefix=always]{changes}
\usepackage[colorlinks=true,linkcolor=blue,urlcolor=blue,citecolor=blue]{hyperref}
\usepackage[toc,page]{appendix}
\usepackage[normalem]{ulem}
\usepackage{orcidlink}
\usepackage{hyphenat}
\usepackage[table]{xcolor}
\usepackage{longtable}
\usepackage{lipsum}
\usepackage{etoolbox}
\usepackage[commandnameprefix=always]{changes}
\usepackage[normalem]{ulem}
\usepackage[tracking=true]{microtype}
\usepackage{lipsum, babel}
\usepackage{sans}
\usepackage{adjustbox}
\usepackage{latexsym}
\usepackage{amsmath}
\usepackage{amssymb}
\usepackage{mathtools}
\usepackage{palatino}
\usepackage{amsfonts}
\usepackage{dcolumn}
\usepackage{bm}
\usepackage{tikz}
\usepackage{bigints}
\usepackage{array,tabularx,multirow,booktabs}
\usepackage[tracking=true]{microtype}
\SetTracking{}{500}
\SetTracking{encoding={*}, shape=sc}{40}
\allowdisplaybreaks
\usepackage{color}
\allowdisplaybreaks
\begin{document} \sloppy

\title{Astrophysical Constraints on Charged Black Holes in Scalar--Tensor--Vector Gravity}

\author{Erdem Sucu
\orcidlink{0009-0000-3619-1492}
}
\email{erdemsc07@gmail.com}
\affiliation{Physics Department, Eastern Mediterranean
University, Famagusta, 99628 North Cyprus, via Mersin 10, T\"{u}rkiye}
\author{K. Boshkayev\orcidlink{http://orcid.org/0000-0002-1385-270X}}
\email{kuantay@mail.ru}
\affiliation{Al-Farabi Kazakh National University, Al-Farabi av. 71, 050040 Almaty, Kazakhstan}
\author{Y. Sekhmani\orcidlink{0000-0001-7448-4579}}
\email{sekhmaniyassine@gmail.com}
\affiliation{Center for Theoretical Physics, Khazar University, 41 Mehseti Street, Baku, AZ1096, Azerbaijan.}
\affiliation{Centre for Research Impact \& Outcome, Chitkara University Institute of Engineering and Technology, Chitkara University, Rajpura, 140401, Punjab, India}
\author{\.{I}zzet Sakall{\i} \orcidlink{0000-0001-7827-9476}}
\email{izzet.sakalli@emu.edu.tr}
\affiliation{Physics Department, Eastern Mediterranean
University, Famagusta, 99628 North Cyprus, via Mersin 10, T\"{u}rkiye}

\author{Mohsen Fathi\orcidlink{0000-0002-1602-0722}}
\email{mohsen.fathi@ucentral.cl}
\affiliation{Centro de Investigaci\'{o}n en Ciencias del Espacio y F\'{i}sica Te\'{o}rica, Universidad Central de Chile, La Serena 1710164, Chile}

\begin{abstract}
We explore charged black holes in Scalar–Tensor–Vector Gravity (STVG), unveiling their distinctive features across multiple physical domains. Our topological analysis reveals that the STVG coupling parameter $\alpha$ bolsters thermal stability while electromagnetic charge $Q$ weakens it. Using the Gauss–Bonnet theorem, we find that $\alpha$ amplifies light deflection and enlarges shadow silhouettes, with $Q$ generating opposite effects. Our quantum-corrected models with exponential entropy terms pinpoint phase transitions in the microscopic regime, modifying conventional thermodynamic relationships. Calculations of strong gravitational lensing, shadow geometry, and Hawking emission show clear STVG signatures that diverge from Einstein's predictions. Notably, our accretion disk analysis uncovers an intriguing phenomenon: specific combinations of $\alpha$ and $Q$ can produce radiation patterns resembling spinning Kerr black holes, creating potential identification challenges for observers. These findings establish concrete observational tests for STVG theory through next-generation astronomical imaging and lensing campaigns. By connecting theoretical predictions to measurable quantities, we outline specific pathways to confirm or constrain STVG using data from current and future space telescopes.
\end{abstract}

\date{\today}

\maketitle

{\color{black}

\section{Introduction} \label{isec1}

The quest to understand gravity's behavior in strong-field environments continues to present significant challenges in theoretical physics. While General Relativity (GR) has been thoroughly validated in weak-field situations, various astrophysical observations—such as anomalies in galactic rotation curves, gravitational lensing, and cosmic expansion—indicate possible deviations from Einstein's theory on larger scales or in more intense gravitational fields \cite{Will:2014kxa,Giddings:2019ujs}. These discrepancies have stimulated the development of modified gravity theories that extend GR while preserving its well-established successes. Among these extensions, STVG, also known as MOdified Gravity (MOG), has emerged as a particularly promising framework \cite{harikumar2022moffat,sucu2025quantumRoyal,izmailov2019modified}.

STVG enhances the gravitational sector by introducing a vector field $\phi_\mu$ alongside a variable gravitational coupling $G = G_N(1+\alpha)$, where $G_N$ represents Newton's constant and $\alpha$ is a dimensionless parameter that controls the deviation from standard GR \cite{harikumar2022moffat}. This additional vector field effectively creates a "gravitational charge" associated with massive objects, modifying gravitational interactions at various scales. STVG has demonstrated considerable success in addressing astrophysical phenomena that challenge conventional dark matter models, including galactic rotation curves, gravitational lensing, and cluster dynamics, without invoking exotic dark matter components \cite{moffat2013mog,moffat2014mog,green2019modified}.

Black holes (BHs)—regions where gravity reaches its extreme limit—provide ideal testing grounds for probing modified gravity theories. The unique properties of charged BHs in STVG theory are particularly valuable, as they encode the intricate interplay between electromagnetic fields and the enhanced gravitational sector \cite{moffat2015black,moffat2021modified}. These solutions exhibit distinctive horizon structures, thermodynamic properties, and optical signatures that potentially differentiate them from their GR counterparts. The parameter $\alpha$, which quantifies the vector field's contribution to gravitational interactions, produces measurable deviations that could be detected through high-precision astrophysical observations \cite{mureika2016black}.

Recent breakthroughs in observational astrophysics have created unprecedented opportunities for testing strong-field gravity. The Event Horizon Telescope (EHT) has captured the first direct images of supermassive BH shadows in M87* and Sagittarius A*, providing detailed measurements of shadow diameters and morphologies \cite{EventHorizonTelescope:2019dse,EventHorizonTelescope:2019uob,EventHorizonTelescope:2019jan,EventHorizonTelescope:2019ths,EventHorizonTelescope:2019pgp,EventHorizonTelescope:2019ggy,EventHorizonTelescope:2022wkp,EventHorizonTelescope:2022xqj}. These images serve not merely as qualitative illustrations but as quantitative benchmarks against which theoretical predictions can be compared. Simultaneously, gravitational wave detectors have opened a new observational window, allowing direct measurement of spacetime dynamics during BH mergers \cite{ligo2022tests,ghosh2017testing}. These complementary observational channels now enable multi-messenger constraints on modified gravity theories in regimes where GR has been traditionally difficult to test \cite{saridakis2023cosmology,baker2021constraining}.

Motivated by these observational advances, our study undertakes an investigation of charged BHs in STVG theory, examining their properties across multiple physical regimes and scales. We analyze both the classical and quantum aspects of these BHs, with particular emphasis on potentially observable signatures that could distinguish STVG from GR in strong-field environments. Our approach integrates analytical techniques from differential geometry, thermodynamics, and relativistic astrophysics to develop a unified understanding of how the STVG parameter $\alpha$ and electromagnetic charge $Q$ influence BH phenomenology.

We begin by analyzing the fundamental structure of charged STVG BHs, deriving their horizon properties and global Hawking temperature using the topological method, which connects thermodynamic quantities to the Euler characteristic of the Euclideanized manifold \cite{robson2019topological,xian2022deriving}. This method demonstrates how $\alpha$ methodically improves thermal stability, while $Q$ tends to reduce it, creating a rich phase structure that differs markedly from Reissner-Nordström BHs in GR.

Next, we investigate gravitational lensing in both vacuum and plasma environments, employing the Gauss-Bonnet theorem (GBT)  \cite{gibbons2008applications,crisnejo2018weak,sucu2025astrophysical,sucu2025probing,sucu2025scalar} to calculate deflection angles for light rays passing near charged STVG BHs. This geometric approach demonstrates that $\alpha$ increases deflection angles while $Q$ reduces them, providing distinctive lensing signatures that could be detected through high-precision astronomical observations. The inclusion of plasma environments further enriches this picture, as the frequency-dependent refractive index creates additional observable effects that could serve as unique tests of the theory \cite{atamurotov2021weak,rahvar2019propagation,sucu2025deflection}.

To extend our analysis beyond the semiclassical regime, we incorporate quantum corrections to BH thermodynamics using an exponential entropy model \cite{sucu2025quantumHassan,gursel2025thermodynamics,sucu2025quantumOzcan,sucu2025nonlinear,WOS:001565141800002}. This approach modifies standard Bekenstein-Hawking relations, introducing corrections to internal energy, free energy, pressure, and the Joule-Thomson coefficient. The resulting phase transitions and critical phenomena provide a window into quantum gravitational effects that might be indirectly probed through astrophysical observations.

Our investigation continues with a detailed analysis of strong gravitational lensing, shadow formation, and energy emission spectra for charged STVG BHs \cite{kuang2022constraining}. These calculations reveal that the parameter $\alpha$ consistently enlarges the photon sphere radius and shadow size, creating observable features that differ from GR predictions. By comparing our theoretical results with EHT measurements, we derive constraints on the STVG parameters, demonstrating how current observations already restrict the viable parameter space of the theory.

Finally, we examine the radiative properties of accretion disks around charged STVG BHs, calculating the flux distribution, temperature profile, and luminosity spectrum using the relativistic thin-disk model \cite{perez2013accretion,pun2008thin,hu2022observational,zheng2025shadows,al2025shadow}. These calculations reveal that variations in $\alpha$ and $Q$ can produce emission characteristics that mimic rotating Kerr BHs in GR, potentially creating observational degeneracies that require multiple complementary measurements to resolve.

The paper is organized as follows. Section \ref{isec2} introduces charged BHs in STVG theory and analyzes their horizon structure and thermodynamic properties. Section \ref{isec3} investigates light deflection in vacuum using the GBT approach. Section \ref{isec4} extends this analysis to plasma environments, revealing frequency-dependent effects. Section \ref{isec5} incorporates quantum corrections to BH thermodynamics, examining phase transitions and Joule-Thomson expansion. Section \ref{isec6} analyzes strong gravitational lensing properties. Section \ref{isec7} investigates BH shadow geometry and compares theoretical predictions with EHT observations. Section \ref{isec8} examines energy emission spectra and Hawking radiation. Section \ref{isec9} explores accretion disk properties and radiative signatures. Finally, Sec. \ref{isec10} discusses the implications of our findings and outlines future research directions.

\section{Charged BH\lowercase{s} in STVG Theory and Their Horizon Structure} \label{isec2}

STVG extends GR by introducing a dynamical vector field $\phi_\mu$ together with a variable effective gravitational coupling \cite{moffat2015black,moffat2006scalar,moffat2013mog}. The theory is parametrized by a dimensionless quantity $\alpha$, which controls deviations from the Einsteinian limit and simultaneously gives rise to a ``gravitational charge'' associated with the vector sector \cite{moffat2021modified}. In the presence of ordinary electromagnetic fields, the natural arena for exploring compact objects becomes that of charged BH solutions.

The starting point is the modified Einstein equations,
\begin{equation}
R_{\mu\nu}-\tfrac{1}{2}g_{\mu\nu}R = T^{(\phi)}_{\mu\nu}+T^{(\text{EM})}_{\mu\nu},
\label{eq:mogfield}
\end{equation}
where $T^{(\phi)}_{\mu\nu}$ corresponds to the stress--energy tensor of the vector field and $T^{(\text{EM})}_{\mu\nu}$ represents the usual Maxwell contribution \cite{Brownstein2009}. For a static and spherically symmetric background, both tensors take the form of an anisotropic fluid with diagonal components proportional to $1/r^{4}$, namely
\begin{equation}
T^{(\phi)\,\mu}{}_{\nu} = \frac{Q_g^2}{8\pi r^{4}}\;\mathrm{diag}(-1,-1,1,1), \label{eq:stress-new1}
\end{equation}
\begin{equation}
T^{(\text{EM})\,\mu}{}_{\nu} = \frac{Q^2}{8\pi r^{4}}\;\mathrm{diag}(-1,-1,1,1),
\label{eq:stress-new2}
\end{equation}
where $Q_g=\sqrt{\alpha}\,M$ is the vector (gravitational) charge associated with the central mass $M$, and $Q$ denotes the electromagnetic charge \cite{kolovs2020quasi}. The effective Newton coupling in this setup becomes $G=(1+\alpha)G_N$.

To capture the spacetime geometry, one assumes the line element
\begin{equation}
ds^2=-f(r)\,dt^2+\frac{dr^2}{f(r)}+r^2(d\theta^2+\sin^2\theta\,d\varphi^2),
\label{eq:metric-new}  
\end{equation}
where $f(r)$ is to be determined from Eq.~\eqref{eq:mogfield}. Substituting the stress--energy expressions into the field equations reduces the problem to a single radial equation,
\begin{equation}
r f'(r)+f(r)-1 - (1+\alpha)\frac{\alpha M^2+Q^2}{r^2}=0.
\label{eq:radialeq-new}
\end{equation}
Its general solution is \cite{nishonov2025qpos}
\begin{equation}
f(r)=1-\frac{2(1+\alpha)M}{r}+(1+\alpha)\frac{\alpha M^2+Q^2}{r^2}.
\label{eq:lapse-new}
\end{equation}

The structure of horizons follows from the condition $f(r)=0$, giving
\begin{equation}
r_{\pm}=(1+\alpha)M\;\pm\;\sqrt{(1+\alpha)M^2 - Q^2}.
\label{eq:horizon-new}
\end{equation}
This expression makes it clear that the parameter $\alpha$ not only rescales the effective gravitational strength but also shifts the location of the horizons relative to the standard RN case \cite{moffat2015black,guo2018observational}. In particular, the extremal configuration is realized when the square root vanishes, leading to
\begin{equation}
|Q|_{\text{ext}}=\sqrt{1+\alpha}M, \qquad r_h^{\text{(min)}}=(1+\alpha)M.
\label{eq:extremal-new}
\end{equation}

\setlength{\tabcolsep}{10pt}
\renewcommand{\arraystretch}{1.3}
\begin{longtable*}{|c|c|c|c|c|}
\hline
\rowcolor{brown!50}
\textbf{$M$} & \textbf{$Q$} & \textbf{$\alpha$} & \textbf{Horizon(s)} & \textbf{Configuration} \\
\hline
\endfirsthead
\hline
\rowcolor{brown!50}
\textbf{$M$} & \textbf{$Q$} & \textbf{$\alpha$} & \textbf{Horizon(s)} & \textbf{Configuration} \\
\hline
\endhead

1.0 & 0.0 & 0.0 & $[2.0]$ & Schwarzchild BH \\
\hline
1.0 & 0.0 & 0.01 & $[2.014987562,\ 0.005012438]$ & Non-extremal BH \\
\hline
1.0 & 0.0 & 0.1 & $[2.148808848,\ 0.051191152]$ & Non-extremal BH \\
\hline
1.0 & 0.0 & 1.0 & $[3.414213562,\ 0.585786438]$ & Non-extremal BH \\
\hline
1.0 & 0.5 & 0.0 & $[1.866025404,\ 0.1339745962]$ & Non-extremal BH \\
\hline
1.0 & 0.5 & 0.01 & $[1.880344759,\ 0.1396552407]$ & Non-extremal BH \\
\hline
1.0 & 0.5 & 0.1 & $[2.008295106,\ 0.1917048938]$ & Non-extremal BH \\
\hline
1.0 & 0.5 & 1.0 & $[3.224744871,\ 0.775255129]$ & Non-extremal BH \\
\hline
1.0 & 1.0 & 0.0 & $[1.0]$ & Extremal BH \\
\hline
1.0 & 1.0 & 0.01 & $[1.010]$ & Extremal BH \\
\hline
1.0 & 1.0 & 0.1 & $[1.10]$ & Extremal BH \\
\hline
1.0 & 1.0 & 1.0 & $[2.0]$ & Extremal BH \\
\hline
1.0 & 1.5 & 0.0 & $[\,]$ & No horizon (naked singularity) \\
\hline
1.0 & 1.5 & 0.01 & $[\,]$ & No horizon (naked singularity) \\
\hline
1.0 & 1.5 & 0.1 & $[\,]$ & No horizon (naked singularity) \\
\hline
1.0 & 1.5 & 1.0 & $[\,]$ & No horizon (naked singularity) \\
\hline

\caption{Horizon structure for different combinations of $M$, $Q$, and $\alpha$. 
The listed values of $r_h$ denote outer and inner horizons when present, revealing transitions between non-extremal, extremal, and naked singular configurations as the charge and coupling parameter $\alpha$ increase.}
\label{ltab:MQalpha3}
\end{longtable*}

Table \ref{ltab:MQalpha3} presents an analysis of horizon structures for different combinations of mass, charge, and coupling parameters. For fixed mass $M=1$, we observe that increasing $\alpha$ while keeping $Q=0$ progressively enlarges both the outer and inner horizons, with the outer horizon growing more rapidly. This demonstrates how the STVG parameter enhances the effective gravitational field, extending the BH's region of influence. When charge is introduced ($Q=0.5$), we see a similar trend, though the horizons are generally smaller than their uncharged counterparts at the same $\alpha$ value, reflecting the repulsive effect of electromagnetic charge. For $Q=1.0$, all configurations become extremal BHs with a single horizon at $r_h = (1+\alpha)M$, in perfect agreement with Eq.~\eqref{eq:extremal-new}. Beyond this critical charge ($Q=1.5$), all configurations become naked singularities regardless of $\alpha$, indicating that the STVG coupling cannot overcome the strong electromagnetic repulsion in this regime.

\begin{figure*}[h!]
  \centering
  \subfigure[Schwarzschild BH ($M=1$, $Q=0$, $\alpha=0$): $r_h=2.0$]{
    \includegraphics[width=0.33\linewidth]{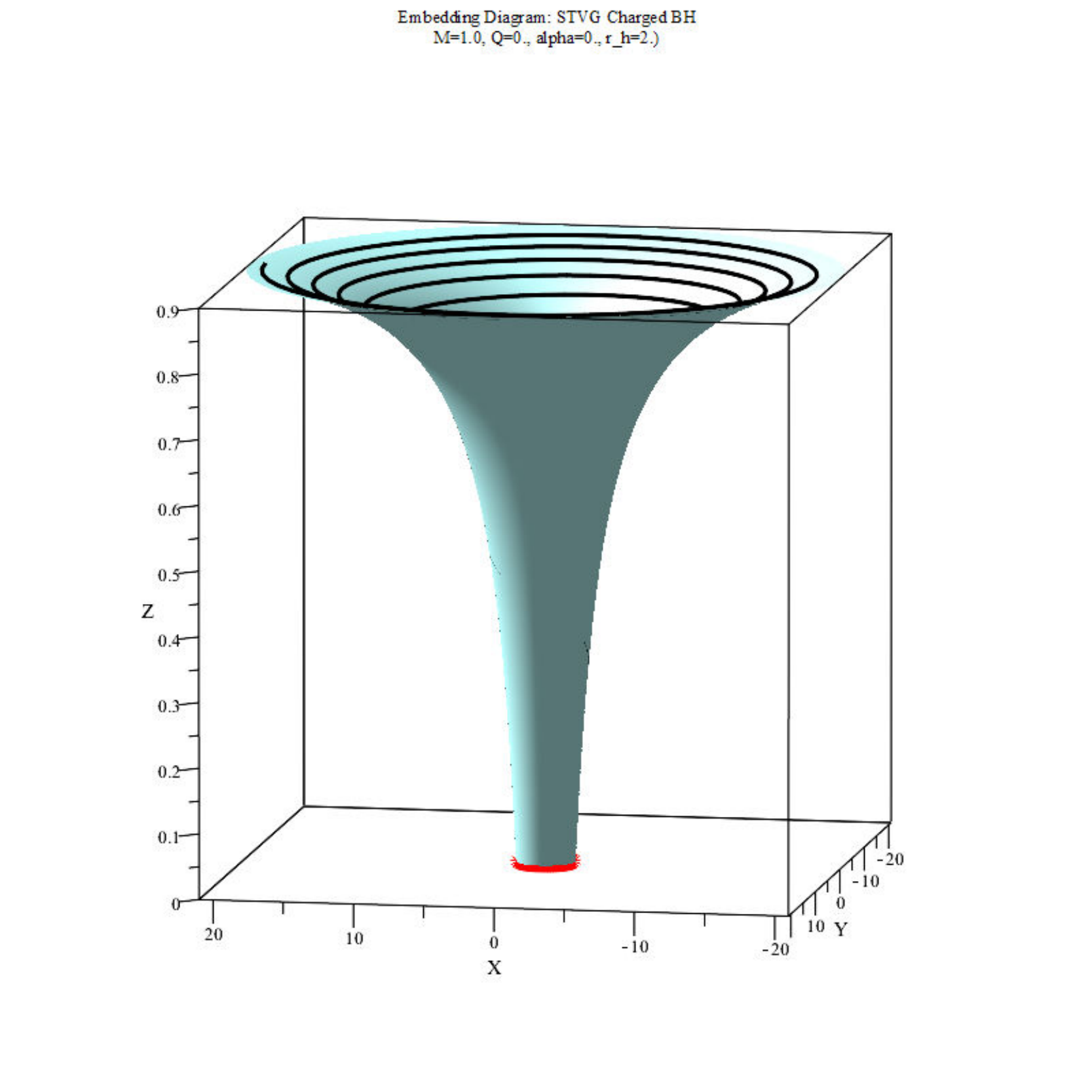}
    \label{fig:3d1}
  }
  \hspace{0.03\linewidth}
  \subfigure[STVG BH ($M=1$, $Q=0$, $\alpha=0.01$): $r_h=2.014987562$]{
    \includegraphics[width=0.33\linewidth]{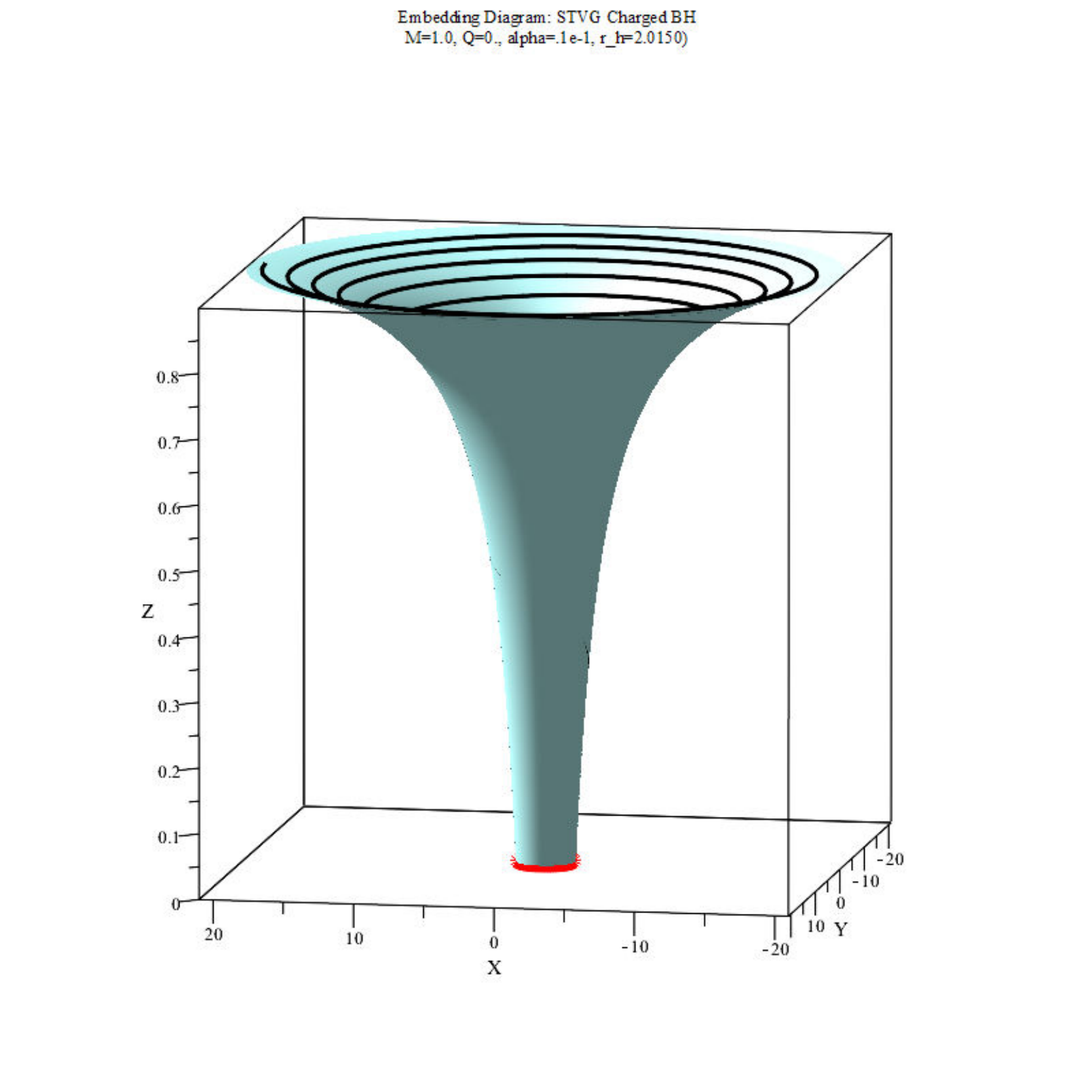}
    \label{fig:3d2}
  }
  \hspace{0.03\linewidth}
  \subfigure[STVG BH ($M=1$, $Q=0$, $\alpha=1.0$): $r_h=3.414213562$]{
    \includegraphics[width=0.33\linewidth]{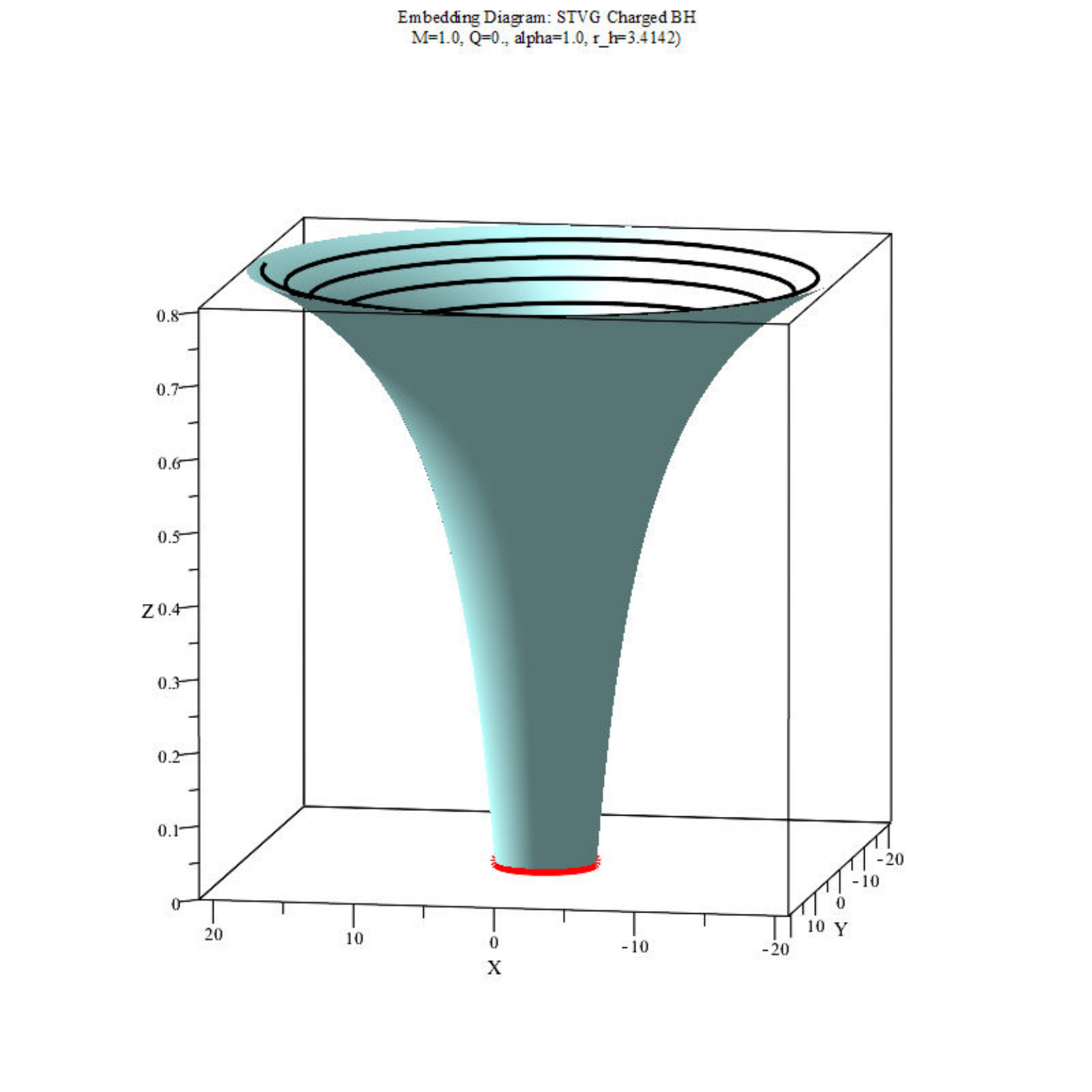}
    \label{fig:3d3}
  }
  \subfigure[STVG Charged BH ($M=1$, $Q=0.5$, $\alpha=1.0$): $r_h=3.224744871$]{
    \includegraphics[width=0.33\linewidth]{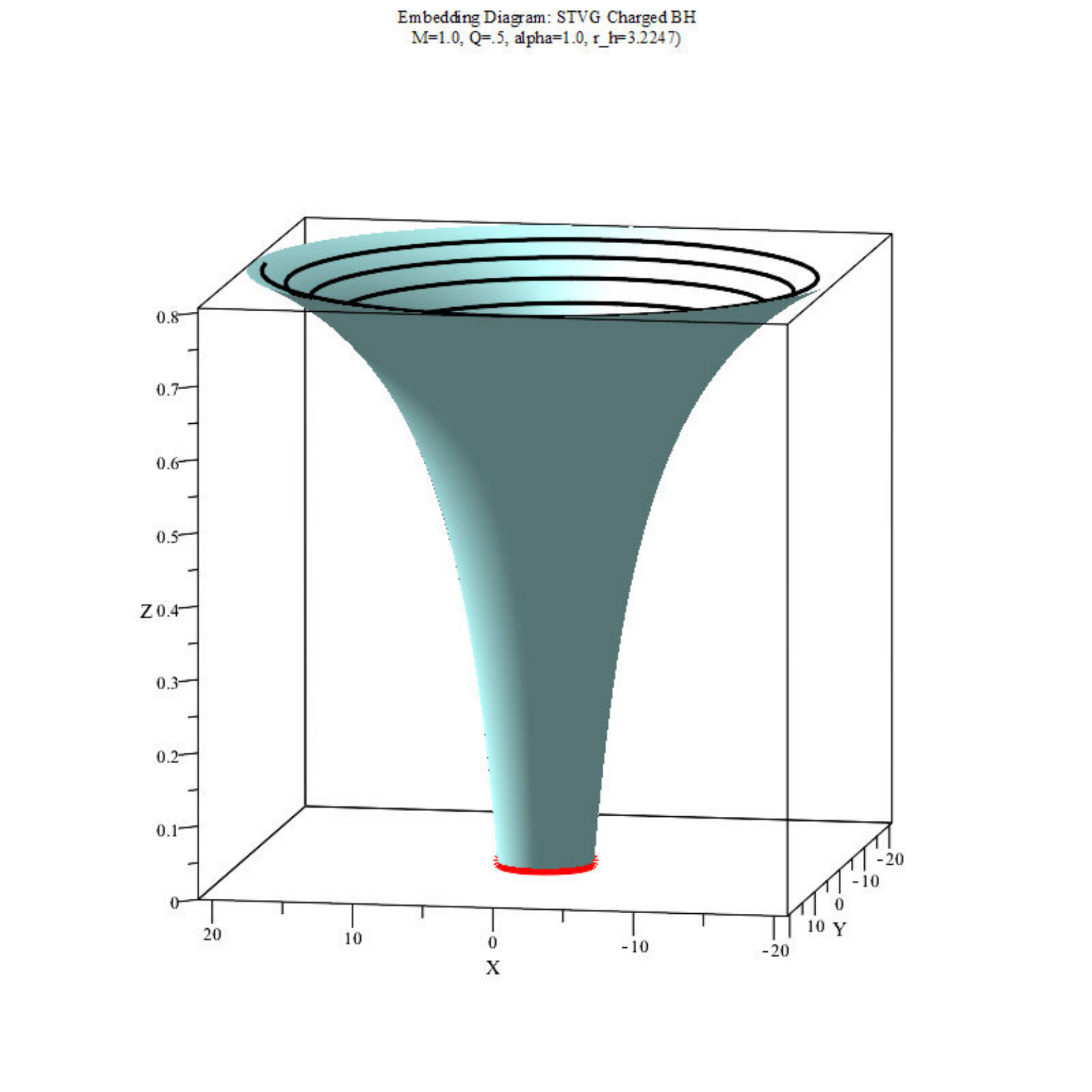}
    \label{fig:3d4}
  }
  \hspace{0.03\linewidth}
  \subfigure[STVG Charged BH ($M=1$, $Q=1$, $\alpha=0.01$): $r_h=1.10$]{
    \includegraphics[width=0.33\linewidth]{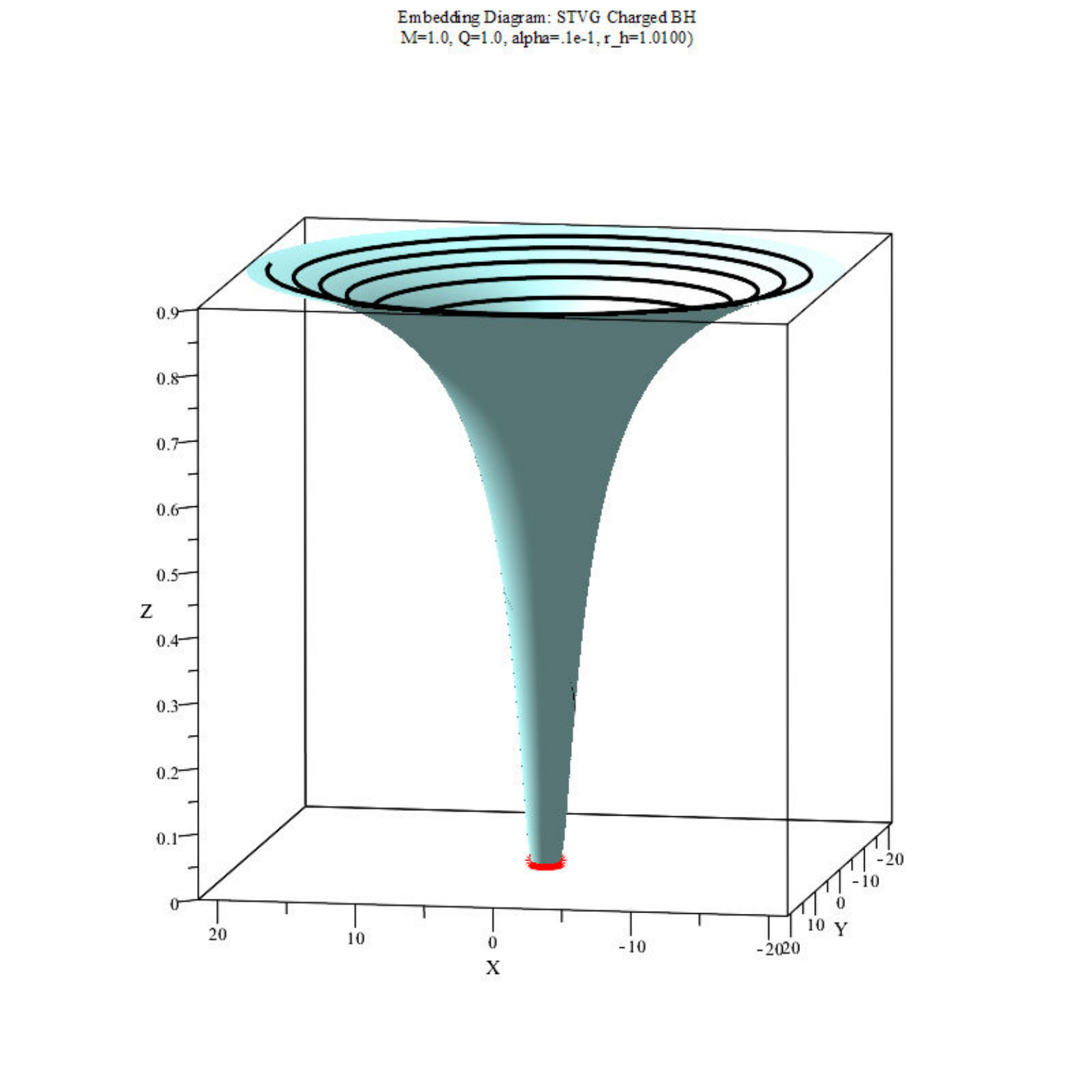}
    \label{fig:3d5}
  }
  \hspace{0.03\linewidth}
  \subfigure[STVG Charged BH ($M=1$, $Q=1$, $\alpha=1.0$): $r_h=2.0$]{
    \includegraphics[width=0.33\linewidth]{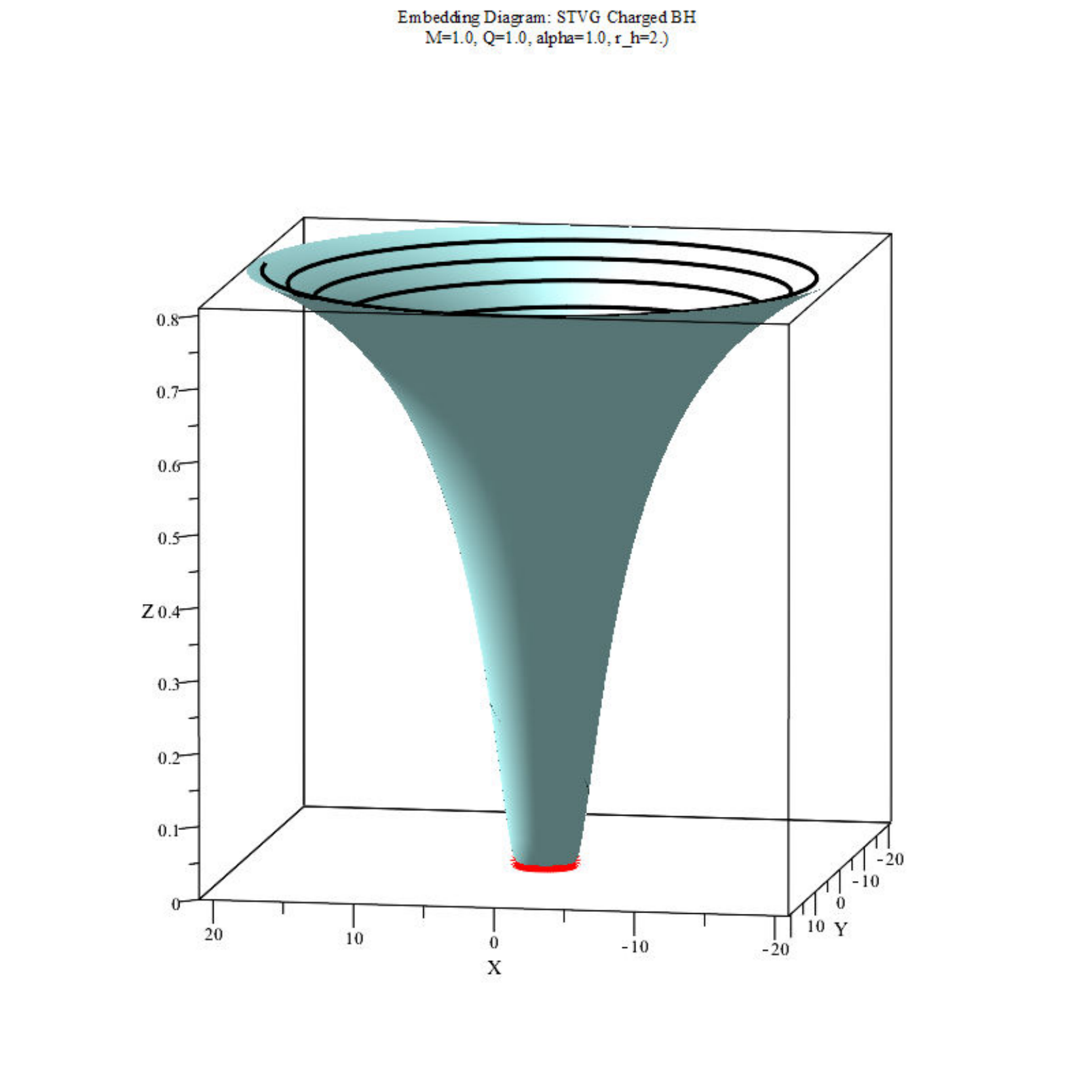}
    \label{fig:3d6}
  }

  \caption{3D embedding diagrams of STVG charged BHs for various combinations of $M$, $Q$, and $\alpha$. 
  \textbf{(a)} The Schwarzschild limit ($Q=0$, $\alpha=0$) establishes the uncharged reference geometry with $r_h=2M$. 
  \textbf{(b–c)} Increasing $\alpha$ in the uncharged case enlarges the horizon radius, illustrating the gravitational enhancement predicted by MOG. 
  \textbf{(d–f)} Charged configurations display the interplay between electromagnetic and scalar–tensor–vector effects: larger $\alpha$ mitigates the electromagnetic repulsion and modifies the curvature profile near the event horizon. 
  The red ring denotes the event horizon, while the black spiral trajectory represents an infalling test particle.}
  \label{fig:stvg3D}
\end{figure*}

\begin{figure}[!ht]
  \centering
\includegraphics[width=0.5\textwidth]{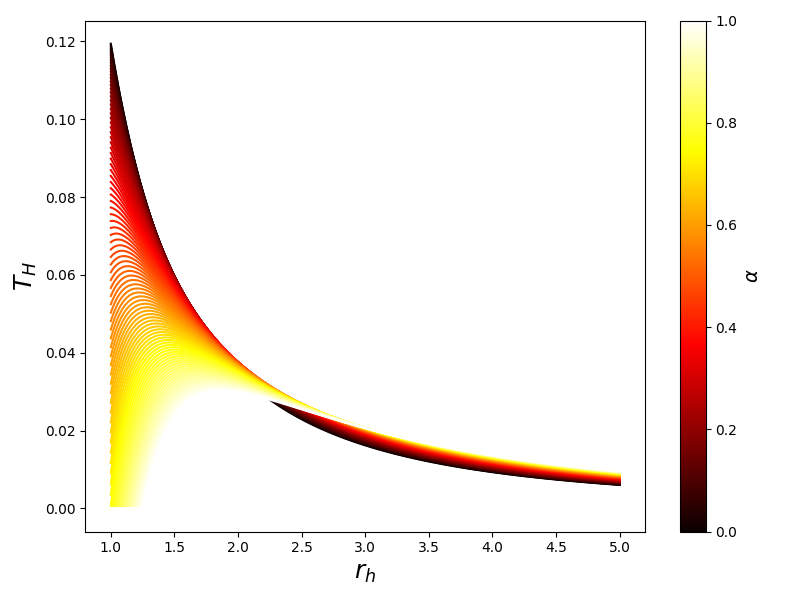}
\caption{Hawking temperature $T_H$ versus horizon radius $r_h$ for different $\alpha$ values with fixed $M=1$ and $Q=0.5$. Increasing $\alpha$ raises the temperature and shifts its peak to larger $r_h$, reflecting stronger effective gravity and modified thermodynamic behavior.}
\label{fig_TH}
\end{figure}

Figure \ref{fig:stvg3D} visualizes the 3D embedding diagrams of STVG charged BHs for various parameter combinations. The subfigures illustrate how the spacetime geometry responds to changes in both $\alpha$ and $Q$. The Schwarzschild case (a) provides the baseline reference with $r_h=2M$. As $\alpha$ increases in the uncharged scenario (b-c), the horizon expands significantly, demonstrating the gravitational enhancement predicted by MOG. The charged configurations (d-f) reveal the competing effects: while the electromagnetic charge tends to contract the horizon and steepen the embedding funnel, larger $\alpha$ values counteract this tendency by strengthening the gravitational attraction. The comparison between subfigures (e) and (f), both with $Q=1$, where increasing $\alpha$ from 0.01 to 1.0 nearly doubles the horizon radius from $r_h=1.10$ to $r_h=2.0$. The red rings marking the event horizons provide clear visual confirmation of these quantitative trends.

\subsection{Topological Derivation of Hawking Temperature}

The topological approach offers a powerful method for computing the Hawking temperature without resorting to the full complexity of a higher-dimensional spacetime \cite{sucu2024effect}. This technique links thermodynamic quantities directly to global topological invariants of the Euclideanized manifold. The global Hawking temperature can be obtained from the Euler characteristic through the expression
\begin{equation}
T_{H} = \frac{\hbar c}{4\pi \chi k_{B}} \sum_{j \leq \chi} \int_{r_{h_j}} \sqrt{|g|} \, \mathcal{R} \, dr,
\label{topoTH}
\end{equation}
where $\mathcal{R}$ denotes the Ricci scalar of the two-dimensional Euclidean section, $g$ is the determinant of the induced metric, and $r_{h_j}$ represents the location of the $j$-th Killing horizon. For STVG charged BHs, we obtain
\begin{equation}
T_{H} =\frac{M}{2 \pi r^{2}}+\frac{M \alpha}{2 \pi r^{2}}-\frac{M^{2} \alpha^{2}}{2 \pi r^{3}}-\frac{\alpha M^{2}}{2 \pi r^{3}}-\frac{Q^{2} \alpha}{2 \pi r^{3}}-\frac{Q^{2}}{2 \pi r^{3}}
\label{eq:hawking-temp}
\end{equation}

Figure \ref{fig_TH} illustrates the behavior of the Hawking temperature as a function of horizon radius for different values of $\alpha$, with $M=1$ and $Q=0.5$. The temperature profile reveals the competition between two terms: the surface gravity contribution scaling as $(1+\alpha)/r_h^{2}$ (positive), and higher-order corrections involving $\alpha^{2}$ and $Q^{2}$ proportional to $1/r_h^{3}$ (negative). For large horizons, $T_H$ follows an inverse-square behavior, with increasing $\alpha$ shifting the temperature upward, indicating an enhancement in effective surface gravity. In the small-horizon regime, the cubic inverse terms dominate, suppressing the temperature more strongly as $\alpha$ increases due to both linear and quadratic contributions. Characteristic peaks in temperature curves mark the transition between these competing effects, with their positions shifting to larger radii as $\alpha$ increases. This demonstrates how the STVG parameter fundamentally alters the thermodynamic behavior of charged BHs by rebalancing the relationship between gravitational attraction and quantum effects near the horizon \cite{ali2025thermal,ko2010modified}.

\section{Light Deflection in STVG Theory via GBT Formalism} \label{isec3}

The bending of light around compact objects serves as one of the most accurate observational probes of the underlying gravitational theory \cite{will2014confrontation,bisnovatyi2010gravitational,molla2022gravitational}. Within the framework of STVG, the effective gravitational interaction is modified by the presence of an additional vector field that enhances the mass term and alters the dynamics of charged BHs. Recent observational progress in astrophysical lensing has achieved sufficient accuracy to detect even small deviations from the predictions of GR, motivating an analysis of photon trajectories in these extended theories \cite{chael2021observing,chen2020new}.

The classical approach to determining the bending angle relies on solving null geodesics, which becomes increasingly complex in modified gravity theories \cite{iorio2009mond,kuang2022constraining}. The GBT, first applied to gravitational lensing by Gibbons and Werner \cite{gibbons2008applications}, reformulates the calculation in terms of the integral of the Gaussian curvature associated with the optical geometry. This method provides a purely geometric interpretation of light deflection and proves particularly efficient in cases where nonlinear couplings or vector-field effects complicate direct geodesic integration \cite{Werner2012}. 

Restricting to the equatorial plane, the null condition yields the two-dimensional optical line element for charged BHs in STVG,
\begin{equation}
dt^2 = \frac{1}{f^{2}(r)}dr^2 + \frac{r^2}{f(r)}d\phi^2,
\label{eq:optical-metric}
\end{equation}
where the lapse function $f(r)$ from Eq.~\eqref{eq:lapse-new} carries information about the mass, charge, and the coupling parameter $\alpha$ of the theory. The corresponding Gaussian curvature of the optical manifold is defined as \cite{sucu2024effect}
\begin{equation}
K = \frac{R_{r\phi r\phi}}{\gamma}=\frac{1}{\sqrt{\gamma}}\left[\frac{\partial}{\partial \phi}\left(\frac{\sqrt{\gamma}}{\gamma_{rr}}\Gamma^{\phi}_{rr}\right)-\frac{\partial}{\partial r}\left(\frac{\sqrt{\gamma}}{\gamma_{rr}}\Gamma^{\phi}_{r\phi}\right)\right],
\label{eq:gaussian-curvature}
\end{equation}
with $R$ denoting the Ricci scalar of the two-dimensional optical space. In the weak-field regime, where the impact parameter $b$ is much larger than the horizon scale, the curvature admits the expansion
\begin{multline}
K \approx \frac{4 M^{4} \alpha^{3}}{r^{6}}+\frac{2 \alpha^{2} M^{4}}{r^{6}}+\frac{2 Q^{4} \alpha^{2}}{r^{6}}+\frac{4 Q^{4} \alpha}{r^{6}}-\frac{2 M \alpha}{r^{3}}\\+\frac{6 M^{2} \alpha^{2}}{r^{4}}+\frac{9 \alpha  M^{2}}{r^{4}}+\frac{3 Q^{2} \alpha}{r^{4}}-\frac{12 M^{3} \alpha^{2}}{r^{5}}\\-\frac{6 M^{3} \alpha}{r^{5}}-\frac{6 M \,Q^{2}}{r^{5}}-\frac{6 M^{3} \alpha^{3}}{r^{5}}+\frac{2 M^{4} \alpha^{4}}{r^{6}}-\frac{2 M}{r^{3}}\\+\frac{3 Q^{2}}{r^{4}}+\frac{3 M^{2}}{r^{4}}+\frac{2 Q^{4}}{r^{6}}-\frac{12 M \,Q^{2} \alpha}{r^{5}}\\-\frac{6 M \alpha^{2} Q^{2}}{r^{5}}+\frac{4 M^{2} \alpha^{3} Q^{2}}{r^{6}}+\frac{8 M^{2} \alpha^{2} Q^{2}}{r^{6}}+\frac{4 \alpha  M^{2} Q^{2}}{r^{6}} + \cdots,
\label{eq:curvature-expansion}
\end{multline}

The GBT formulation connects this curvature to the deflection angle through the surface integral
\begin{equation}
\hat{\alpha} = - \lim_{R \to \infty} \int_0^{\pi} \int_{b/\sin\phi}^{R} K \sqrt{\det g}\, dr d\phi,
\label{eq:gbt-integral}
\end{equation}
with $\sqrt{\det g}$ representing the determinant of the optical metric. This global geometric relation emphasizes that light bending depends on the integrated curvature of the effective optical space rather than the local geodesic structure alone \cite{Perlick2004,ono2018deflection}. Evaluating the integral with the curvature expansion leads to the approximate bending angle
\begin{multline}
\Theta \simeq \frac{27 M^{2} Q^{2} \pi}{16 b^{4}}-\frac{3 M^{4} \pi  \,\alpha^{4}}{16 b^{4}}+\frac{21 M^{4} \alpha^{3} \pi}{16 b^{4}}\\+\frac{51 M^{4} \alpha^{2} \pi}{16 b^{4}}-\frac{3 M^{2} \pi  \,\alpha^{2}}{2 b^{2}}-\frac{3 \pi  Q^{4} \alpha^{2}}{16 b^{4}}+\frac{27 M^{4} \alpha  \pi}{16 b^{4}}-\frac{3 M^{2} \pi  \alpha}{4 b^{2}}\\-\frac{64 \alpha  M^{3} Q^{2}}{25 b^{5}}-\frac{64 Q^{4} \alpha  M}{25 b^{5}}+\frac{4 M \,Q^{2} \alpha}{3 b^{3}}-\frac{3 Q^{4} \alpha  \pi}{8 b^{4}}\\-\frac{3 \pi  Q^{2} \alpha}{4 b^{2}}-\frac{64 M^{3} \alpha^{3} Q^{2}}{25 b^{5}}-\frac{128 M^{3} \alpha^{2} Q^{2}}{25 b^{5}}-\frac{32 Q^{4} \alpha^{2} M}{25 b^{5}}\\+\frac{8 M \,Q^{2} \alpha^{2}}{3 b^{3}}-\frac{4 M^{3}}{b^{3}}+\frac{4 M}{b}-\frac{3 M^{2} \pi  Q^{2} \alpha^{3}}{8 b^{4}}\\+\frac{15 M^{2} Q^{2} \alpha^{2} \pi}{16 b^{4}}+\frac{3 M^{2} Q^{2} \alpha  \pi}{b^{4}}-\frac{32 M^{5} \alpha^{4}}{25 b^{5}}-\frac{64 M^{5} \alpha^{3}}{25 b^{5}}\\+\frac{8 M^{3} \alpha^{3}}{3 b^{3}}-\frac{32 \alpha^{2} M^{5}}{25 b^{5}}-\frac{8 M^{3} \alpha^{2}}{3 b^{3}}-\frac{28 M^{3} \alpha}{3 b^{3}}+\frac{4 M \alpha}{b}\\-\frac{32 Q^{4} M}{25 b^{5}}-\frac{4 M \,Q^{2}}{3 b^{3}}+\frac{3 M^{2} \pi}{4 b^{2}}-\frac{3 Q^{4} \pi}{16 b^{4}}-\frac{3 \pi  Q^{2}}{4 b^{2}}.
\label{eq:bending-angle}
\end{multline}

Several important features emerge from this expression. First, in the limit $\alpha \to 0$ and $Q \to 0$, we recover the standard Schwarzschild result $\Theta \approx 4M/b$, confirming the consistency of our approach \cite{arakida2012effect,Virbhadra2000}. Second, the terms containing $\alpha$ contribute positively to the deflection angle, while those with $Q$ tend to reduce it, reflecting the competing effects of the gravitational and electromagnetic interactions. Third, the deflection angle contains terms with different power-law dependencies on $b$, ranging from $b^{-1}$ to $b^{-5}$, demonstrating the rich phenomenology of STVG compared to GR.

Figure \ref{fig_lens} illustrates the variation of the deflection angle $\Theta$ for light rays propagating in the background of a charged BH within STVG, evaluated via the GBT. For the chosen parameters $M=1$ and $Q=0.5$, the figure reveals how the coupling parameter $\alpha$ directly influences the curvature of spacetime and consequently the bending of light. As $\alpha$ increases, the deflection angle increases significantly, indicating that the vector-tensor interaction strengthens the effective gravitational attraction and enhances the curvature of the optical manifold. In contrast, for smaller or vanishing $\alpha$, the results smoothly approach the RN limit of GR, where the electric charge slightly weakens the deflection compared to the Schwarzschild case.

This result complements the BH horizon analysis presented in Sec. \ref{isec2} and Table \ref{ltab:MQalpha3}, where we observed that increasing $\alpha$ enlarges the horizon radius. Both effects stem from the same underlying mechanism: the STVG parameter enhances the effective gravitational field, leading to both stronger light bending and expanded horizon structure. The 3D diagrams in Fig. \ref{fig:stvg3D} provide visual confirmation of how the spacetime geometry is modified by $\alpha$, which directly translates to the lensing properties analyzed here.

These findings suggest that precise observations of light bending around compact objects could serve as a sensitive probe of deviations from GR. By measuring the deflection angles with sufficient accuracy, one could potentially constrain the STVG parameter $\alpha$ and distinguish between competing gravitational theories \cite{kuang2022constraining,silverman1980gravitational}. 

\begin{figure}[!ht]
  \centering
\includegraphics[width=0.5\textwidth]{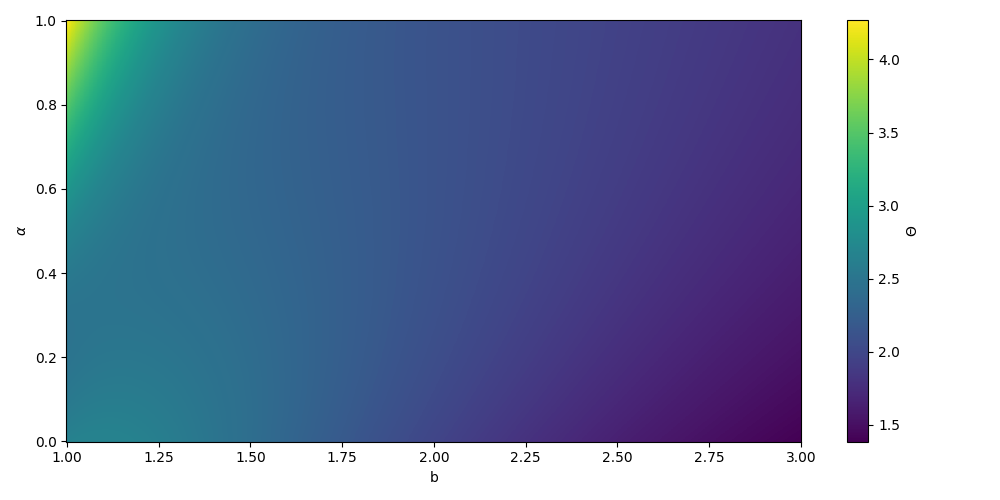}
\caption{Deflection angle $\Theta$ versus impact parameter $b$ for $M=1$ and $Q=0.5$. Increasing $\alpha$ enhances spacetime curvature and light bending, while $\alpha \to 0$ recovers the RN limit of GR.}
\label{fig_lens}
\end{figure}

\section{Gravitational Lensing by STVG BHs in Plasma Environments} \label{isec4}

Extending our analysis from Sec. \ref{isec3}, we now investigate how plasma environments influence the gravitational lensing properties of charged BHs in STVG theory. The presence of plasma introduces frequency-dependent dispersive effects that modify the propagation of electromagnetic waves, resulting in a more complex lensing phenomenon than in vacuum \cite{bisnovatyi2010gravitational,BisnovatyiKoganTsupko2010}. This interplay between plasma dispersion and modified gravity creates distinctive signatures that could potentially be detected in astrophysical observations \cite{sucu2025charged}.

In plasma environments, light propagation is characterized by a refractive index $n$ that depends on both the spatial coordinate and the frequency of radiation \cite{perlick2015influence,sucu2025exploring}. For a cold, non-magnetized plasma, this refractive index takes the form
\begin{equation}
n^2(r, \varpi(r)) = 1 - \frac{\varpi_e^2(r)}{\varpi_\infty^2(r)},
\label{eq:refractive-index1}
\end{equation}
where $\varpi_e(r)$ represents the plasma frequency and $\varpi_\infty(r)$ denotes the photon frequency measured by a distant observer \cite{rogers2018gravitational,rogers2015frequency}. Using the STVG metric function $f(r)$ from Eq.~\eqref{eq:lapse-new}, the refractive index can be expressed as
\begin{equation}
n(r) = \sqrt{1 - \frac{\varpi_e^2}{\varpi_\infty^2} f(r)}.
\label{eq:refractive-index2}
\end{equation}

The presence of plasma modifies the effective optical geometry experienced by light rays. Following the approach developed in Sec. \ref{isec3}, we obtain the corresponding optical metric
\begin{equation}
d\tau^2 = g_{opt}^{xy} dx_x dx_y = n^2 \left(\frac{dr^2}{f^2(r)} + \frac{r^2 d\phi^2}{f(r)}\right),
\label{eq:plasma-optical-metric}
\end{equation}
which differs from the vacuum case in Eq.~\eqref{eq:optical-metric} by the multiplication with the square of the refractive index \cite{crisnejo2018weak,sucu2025exploring}.

The Gaussian optical curvature, central to the GBT approach, can be expressed in terms of the curvature tensor as
\begin{equation}
K = \frac{R_{r\phi r\phi}(g_{opt}^{xy})}{\det(g_{opt}^{xy})},
\label{eq:plasma-curvature}
\end{equation}
which in the weak-field limit expands to:
\begin{multline}
K \approx \frac{12 M \,Q^{2} \alpha}{r^{5}}-\frac{6 M \,Q^{2} \alpha^{2}}{r^{5}}-\frac{62 M^{3} \alpha  \varpi_{e}^{2}}{\varpi_{\infty}^{2} r^{5}}-\frac{88 M^{3} \alpha^{2} \varpi_{e}^{2}}{\varpi_{\infty}^{2} r^{5}}\\+\frac{29 M^{2} \alpha  \varpi_{e}^{2}}{\varpi_{\infty}^{2} r^{4}}-\frac{26 M \,Q^{2} \varpi_{e}^{2}}{\varpi_{\infty}^{2} r^{5}}-\frac{38 M^{3} \alpha^{3} \varpi_{e}^{2}}{\varpi_{\infty}^{2} r^{5}}+\frac{17 M^{2} \alpha^{2} \varpi_{e}^{2}}{\varpi_{\infty}^{2} r^{4}}\\-\frac{3 \alpha  M \varpi_{e}^{2}}{\varpi_{\infty}^{2} r^{3}}+\frac{5 \alpha  Q^{2} \varpi_{e}^{2}}{\varpi_{\infty}^{2} r^{4}}-\frac{6 M^{3} \alpha}{r^{5}}-\frac{12 M^{3} \alpha^{2}}{r^{5}}\\+\frac{9 M^{2} \alpha}{r^{4}}-\frac{6 M \,Q^{2}}{r^{5}}-\frac{6 M^{3} \alpha^{3}}{r^{5}}+\frac{6 M^{2} \alpha^{2}}{r^{4}}-\frac{2 \alpha  M}{r^{3}}\\+\frac{3 \alpha  Q^{2}}{r^{4}}-\frac{12 M^{3} \varpi_{e}^{2}}{\varpi_{\infty}^{2} r^{5}}+\frac{12 M^{2} \varpi_{e}^{2}}{\varpi_{\infty}^{2} r^{4}}-\frac{3 M \varpi_{e}^{2}}{\varpi_{\infty}^{2} r^{3}}\\+\frac{5 Q^{2} \varpi_{e}^{2}}{\varpi_{\infty}^{2} r^{4}}+\frac{3 M^{2}}{r^{4}}-\frac{2 M}{r^{3}}+\frac{3 Q^{2}}{r^{4}}\\-\frac{52 M \,Q^{2} \alpha  \varpi_{e}^{2}}{\varpi_{\infty}^{2} r^{5}}-\frac{26 M \,Q^{2} \alpha^{2} \varpi_{e}^{2}}{\varpi_{\infty}^{2} r^{5}}.
\label{eq:plasma-curvature-expansion}
\end{multline}

To calculate the deflection angle in the weak-field regime, we employ the GBT with a straight-line approximation, setting $r = \frac{b}{\sin \phi}$ in the integral
\begin{equation}
\beta = -\lim_{R \to 0} \int_0^\pi \int_{\frac{b}{\sin \phi}}^\infty K \, dS.
\label{eq:plasma-gbt-integral}
\end{equation}

This yields the plasma-modified deflection angle $\beta$, which depends on the impact parameter $b$, the STVG coupling $\alpha$, the charge $Q$, and the plasma parameter $\delta = \frac{\varpi_e^2(r)}{\varpi_\infty^2(r)}$:
\begin{multline}
    \beta \approx \frac{117 M^{2} \pi  Q^{2} \alpha  \delta}{8  b^{4}}+\frac{117 M^{2} \pi  Q^{2} \alpha^{2} \delta}{16  b^{4}}+\frac{3 M^{2} \pi}{4 b^{2}}-\frac{3 \pi  Q^{2}}{4 b^{2}}\\+\frac{27 M^{4} \pi  \delta}{8  b^{4}}-\frac{3 M^{2} \pi  \delta}{4  b^{2}}-\frac{5 \pi  Q^{2} \delta}{4  b^{2}}+\frac{27 M^{2} \pi  Q^{2} \alpha^{2}}{16 b^{4}}\\+\frac{27 M^{2} \pi  Q^{2} \alpha}{8 b^{4}}+\frac{4 \alpha  M}{b}+\frac{171 M^{4} \pi  \,\alpha^{3} \delta}{16  b^{4}}+\frac{99 M^{4} \pi  \,\alpha^{2} \delta}{4  b^{4}}\\-\frac{17 M^{2} \pi  \,\alpha^{2} \delta}{4  b^{2}}+\frac{279 M^{4} \pi  \alpha  \delta}{16  b^{4}}-\frac{5 M^{2} \pi  \alpha  \delta}{ b^{2}}\\-\frac{5 \pi  Q^{2} \alpha  \delta}{4  b^{2}}+\frac{117 M^{2} \pi  Q^{2} \delta}{16  b^{4}}+\frac{152 M^{3} \alpha^{3} \delta}{9  b^{3}}\\+\frac{4 M}{b}+\frac{6 M \delta}{ b}+\frac{8 M \,Q^{2} \alpha^{2}}{3 b^{3}}+\frac{6 M \delta \alpha}{ b}\\+\frac{104 M \,Q^{2} \alpha^{2} \delta}{9  b^{3}}+\frac{148 \alpha  Q^{2} \delta M}{9  b^{3}}-\frac{4 M \,Q^{2}}{3 b^{3}}+\frac{27 M^{4} \pi  \,\alpha^{3}}{16 b^{4}}\\+\frac{27 M^{4} \pi  \,\alpha^{2}}{8 b^{4}}-\frac{3 M^{2} \pi  \,\alpha^{2}}{2 b^{2}}+\frac{27 M^{4} \pi  \alpha}{16 b^{4}}-\frac{3 M^{2} \pi  \alpha}{4 b^{2}}\\-\frac{3 \pi  Q^{2} \alpha}{4 b^{2}}+\frac{27 M^{2} \pi  Q^{2}}{16 b^{4}}+\frac{148 M^{3} \alpha^{2} \delta}{9  b^{3}}\\-\frac{32 M^{3} \delta}{3  b^{3}}-\frac{28 M^{3} \alpha}{3 b^{3}}-\frac{8 M^{3} \alpha^{2}}{3 b^{3}}+\frac{4 M \,Q^{2} \alpha}{3 b^{3}}+\frac{44 M \,Q^{2} \delta}{9  b^{3}}\\+\frac{8 M^{3} \alpha^{3}}{3 b^{3}}-\frac{4 M^{3}}{b^{3}}-\frac{100 M^{3} \alpha  \delta}{9  b^{3}}
\label{eq:plasma-deflection-angle}
\end{multline}

Several notable features emerge from this expression. First, the leading terms in the absence of plasma ($\delta = 0$) reduce to those derived in Eq.~\eqref{eq:bending-angle} for vacuum lensing. Second, the plasma parameter $\delta$ introduces additional terms that couple with the STVG parameter $\alpha$ and the charge $Q$, creating a richer phenomenology than in either standard GR or vacuum STVG \cite{crisnejo2018weak,kala2025gravitational}. Third, unlike in GR, where plasma generally reduces the deflection angle, in STVG the interaction between $\alpha$ and $\delta$ can enhance the deflection for certain parameter combinations \cite{tsupko2013gravitational}.

Figure \ref{fig:beta_density} illustrates the density distribution of the deflection angle $\beta(b,\delta)$ for different values of the STVG coupling parameter $\alpha = (0, 0.5, 1)$ in a plasma medium surrounding a charged BH. The three panels clearly demonstrate how increasing $\alpha$ progressively enhances the deflection angle across the entire parameter space of impact parameters $b$ and plasma densities $\delta$. This enhancement appears as an expansion and intensification of the high-$\beta$ region (brighter colors) in the density plots.

The figure reveals several important physical insights that complement our earlier analysis of horizon structure in Table \ref{ltab:MQalpha3} and 3D diagrams in Fig. \ref{fig:stvg3D}. First, just as $\alpha$ enlarges the horizon radius and intensifies spacetime curvature near the BH, it similarly strengthens the lensing effect at larger distances. Second, the electromagnetic charge $Q$ (fixed at $Q = 0.5M$ in this figure) counteracts the gravitational attraction, reducing $\beta$ especially at small impact parameters. This creates a clear competition between $\alpha$ and $Q$: while $\alpha$ amplifies gravitational bending, $Q$ suppresses it through electrostatic repulsion.

An additional important feature visible in Fig. \ref{fig:beta_density} is the influence of the plasma parameter $\delta$. Increasing $\delta$ enhances the bending angle due to the refractive delay caused by the plasma medium. This effect becomes more pronounced as $\alpha$ increases, creating a synergistic amplification of the deflection angle when both $\alpha$ and $\delta$ are large. This distinctive coupling between modified gravity and plasma effects provides a potential observational signature that could be detected in radio astronomical observations, where frequency-dependent lensing can be measured with high precision.

These results extend the vacuum analysis presented in Sec. \ref{isec3} and demonstrate that plasma environments offer additional opportunities to test and constrain STVG theory through gravitational lensing observations. The significant strengthening of light deflection caused by the STVG parameter $\alpha$, especially in plasma-rich environments, suggests that radio observations of lensed sources may provide a promising avenue for detecting deviations from GR and constraining alternative gravitational theories \cite{wang2017shadow}.

\begin{figure}[htbp]
    \centering
    \includegraphics[width=0.5\textwidth]{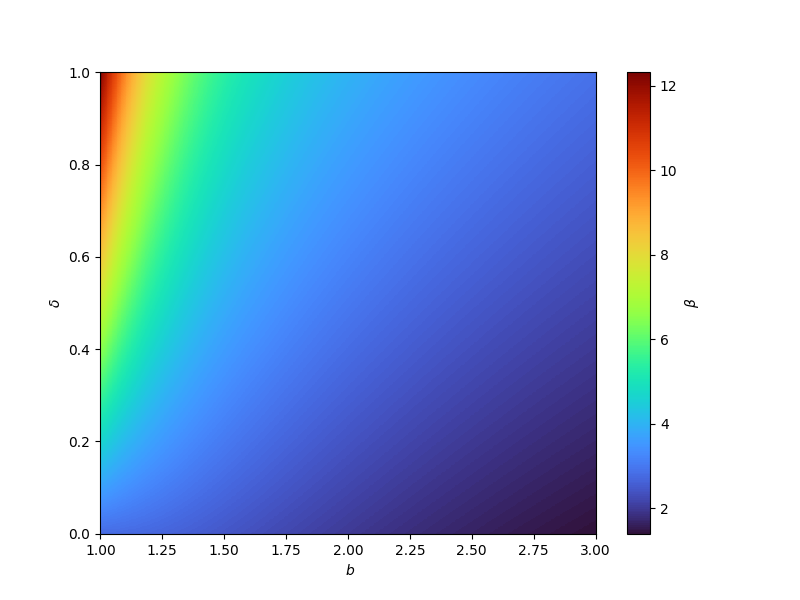}
    \vspace{0.5cm}

    \includegraphics[width=0.5\textwidth]{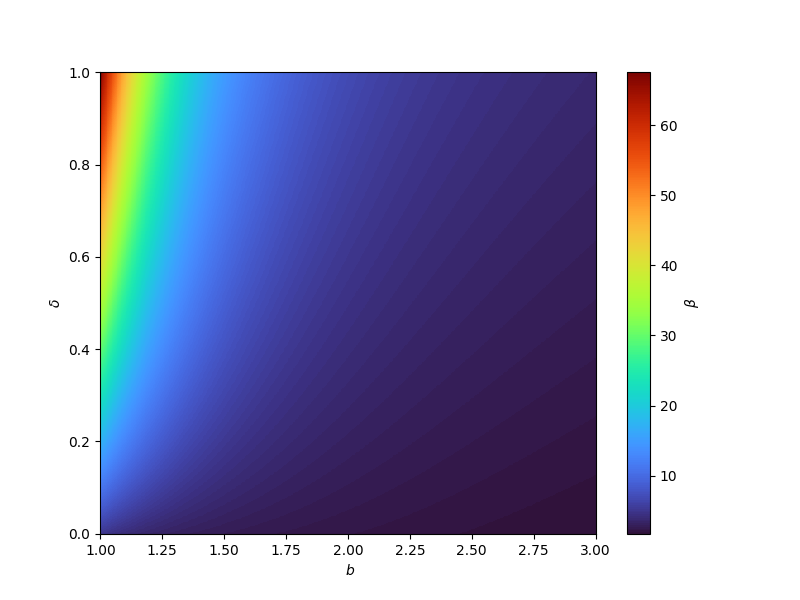}
    \vspace{0.5cm}
    
    \includegraphics[width=0.5\textwidth]{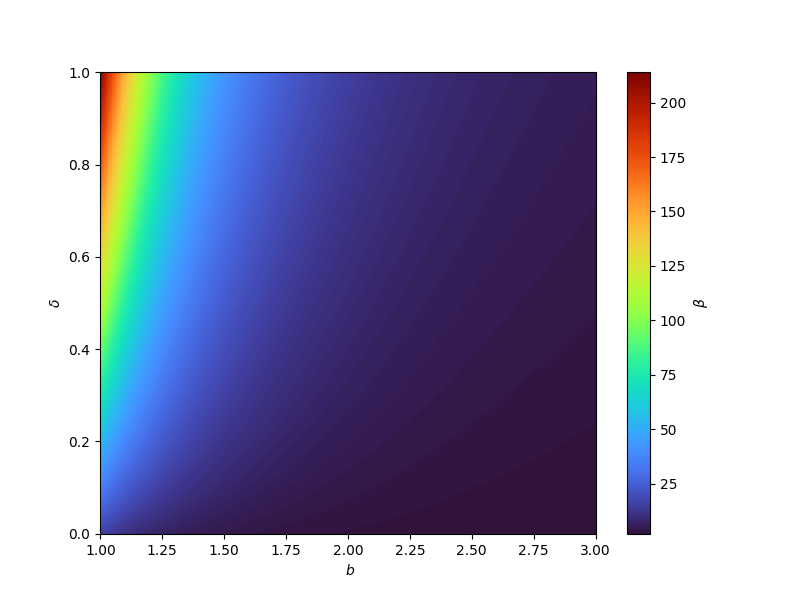}
    
    \caption{Comparison of $\beta(b,\delta)$ density plots for different parameter choices $\alpha$ (0,0.5,1) respectively. The color intensity indicates deflection angle magnitude, showing how increasing $\alpha$ strengthens gravitational lensing across all impact parameters and plasma densities.}
    \label{fig:beta_density}
\end{figure}

\section{Quantum-Corrected Thermodynamics and JTE Analysis of STVG BHs} \label{isec5}

As BHs undergo Hawking evaporation, their thermodynamic properties require a quantum-corrected framework, particularly when their dimensions approach the Planck scale \cite{AdlerChenSantiago2001,scardigli1999generalized,gecim2019quantum,gecim2020quantum,tekincay2021exotic,tekincay2021zitterbewegung}. At these extremes, microstate fluctuations and quantum effects significantly alter classical thermodynamic relations \cite{dabholkar2005exact,krasnov1999quantum}. The Bekenstein-Hawking entropy, traditionally tied to the event horizon's surface area, receives quantum corrections derived from microstate counting in statistical mechanics \cite{bekenstein2008bekenstein}. These corrections account for quantum fluctuations under fixed energy and particle constraints, introducing subleading terms to the entropy formula.

Among various correction forms proposed in the literature, including logarithmic and power-law terms, an exponential correction has emerged in certain microstate models \cite{chatterjee2020exponential,gursel2025thermodynamics}. In this approach, the entropy is adjusted by a rapidly decaying term proportional to $e^{-S_0}$, yielding the quantum-corrected entropy \cite{sucu2025quantumHassan}:

\begin{equation}
S = S_0 + e^{-S_0}, \label{eq:entropy_corrected}
\end{equation}

where $S_0 = \pi r_h^2$ is the classical Bekenstein-Hawking entropy \cite{ali2024quantum,sucu2025exploring}. This exponential form offers several advantages over alternative corrections: it emerges naturally from microstate counting, decays rapidly for large $S_0$ (ensuring minimal impact on macroscopic BHs), and preserves thermodynamic stability \cite{nozari2012minimal,nozari2008hawking,banerjee2023revisiting}. It provides finite corrections to thermodynamic quantities and aligns with results from both loop quantum gravity and string theory models \cite{banerjee2008quantum,gecim2017gup}.

Building upon our thermodynamic analysis in Sec. \ref{isec2}, we now investigate how these quantum corrections modify the thermodynamic behavior of charged STVG BHs. The quantum-corrected internal energy is derived from the fundamental relation $dE = T_H dS$ with vanishing pressure ($P = 0$). Using Eq.~\eqref{eq:entropy_corrected} and the Hawking temperature from Eq.~\eqref{eq:hawking-temp}, we obtain:

\begin{equation}
dE_C = T_H \left( 1 - e^{-S_0} \right) dS_0,
\label{eq:energy_differential}
\end{equation}

Integrating with $S_0 = \pi r_h^2$ yields:

\begin{multline}
E_C \approx -\pi  \alpha  M^{2} r+\frac{1}{2} \pi  M \,r^{2}-\pi  Q^{2} r-\pi  \,\alpha^{2} M^{2} r\\+\frac{1}{2} \pi  \alpha  M \,r^{2}-\pi  \alpha  Q^{2} r \label{eq:energy_corrected}
\end{multline}

Figure \ref{fig_enerji} illustrates the variation of $E_C$ with respect to $r_h$ for fixed parameters $M = 1$ and $Q = 0.5$, and for different values of $\alpha$. The behavior reveals a clear transition between quantum and classical regimes: at small $r_h$, $E_C$ assumes negative values, indicating a quantum-dominated phase where vacuum fluctuations and microstate corrections outweigh classical contributions. As $r_h$ increases, $E_C$ becomes positive and grows monotonically, reflecting the gradual recovery of classical thermodynamic behavior. This transition connects directly to the BH stability analysis in Table \ref{ltab:MQalpha3}, where we observed how varying $\alpha$ and $Q$ affects the horizon structure.

The coupling parameter $\alpha$ plays a decisive role in this evolution: increasing $\alpha$ shifts the $E_C$ curve upward and enhances the internal energy, signifying that the vector-tensor coupling strengthens the gravitational self-energy and stabilizes the configuration. This effect complements our findings in Secs. \ref{isec3} and \ref{isec4}, where we observed that $\alpha$ enhances both the light deflection angle and the BH shadow size. All these phenomena stem from the same physical mechanism: $\alpha$ intensifies the effective gravitational field, thereby strengthening both the spacetime curvature (affecting light propagation) and the system's internal energy (affecting thermodynamic stability).

To further analyze the thermodynamic properties, we compute the Helmholtz free energy:

\begin{equation}
F_C = -\int S dT_H, \label{eq:helmholtz}
\end{equation}

which yields:

\begin{multline}
F_C \approx  -\frac{3 \pi  \alpha  M^{2} r}{4}+\frac{\pi  M \,r^{2}}{4}-\frac{3 \pi  Q^{2} r}{4}+\frac{\alpha  M^{2}}{2 \pi  r^{3}}\\+\frac{Q^{2}}{2 \pi  r^{3}}-\frac{M}{2 \pi  r^{2}}-\frac{3 \pi  \,\alpha^{2} M^{2} r}{4}+\frac{\pi  \alpha  M \,r^{2}}{4}\\-\frac{3 \pi  \alpha  Q^{2} r}{4}+\frac{M^{2} \alpha^{2}}{2 \pi  r^{3}}+\frac{Q^{2} \alpha}{2 \pi  r^{3}}-\frac{M \alpha}{2 \pi  r^{2}} \label{eq:helmholtz_corrected}
\end{multline}

Figure \ref{fig_FF} shows the variation of $F_C$ as a function of $r_h$ for different values of $\alpha$. For small $r_h$, $F_C$ takes large negative values, corresponding to a strongly bound but thermodynamically unstable configuration dominated by quantum effects. As $r_h$ increases, $F_C$ rises smoothly and approaches zero from below, indicating a transition toward a more stable regime. The presence of local minima in the $F_C(r_h)$ curves signals the existence of metastable equilibrium states, reminiscent of first-order phase transitions in Van der Waals systems \cite{kubizvnak2017black,sahan2025quantum}. Increasing $\alpha$ lowers the free energy magnitude at large radii, enhancing global stability and reducing the likelihood of phase coexistence.

The quantum-corrected pressure is determined from:

\begin{equation}
P_C = -\frac{d F_C}{d V}, \label{eq:pressure}
\end{equation}

which gives:

\begin{multline}
P_C \approx  \frac{3 M^{2} \alpha}{16 r^{2}}-\frac{M}{8 r}+\frac{3 Q^{2}}{16 r^{2}}+\frac{3 \alpha  M^{2}}{8 \pi^{2} r^{6}}\\+\frac{3 Q^{2}}{8 \pi^{2} r^{6}}-\frac{M}{4 \pi^{2} r^{5}}+\frac{3 M^{2} \alpha^{2}}{16 r^{2}}-\frac{\alpha  M}{8 r}\\+\frac{3 Q^{2} \alpha}{16 r^{2}}+\frac{3 M^{2} \alpha^{2}}{8 \pi^{2} r^{6}}+\frac{3 Q^{2} \alpha}{8 \pi^{2} r^{6}}-\frac{M \alpha}{4 \pi^{2} r^{5}} \label{eq:pressure_corrected}
\end{multline}

Figure \ref{fig_pp} shows $P_C$ as a function of $r_h$ for three representative values of $\alpha$. Initially, $P_C$ takes negative values for small $r_h$, indicating a tension-like regime with inward pressure. As $r_h$ increases, $P_C$ turns positive, marking a transition to a stable phase where effective pressure balances gravitational attraction and thermal radiation. The crossover point between negative and positive regions represents a critical radius at the onset of mechanical stability \cite{chabab2017phase}. Higher $\alpha$ values shift this transition toward smaller $r_h$ and increase $P_C$ in the large-radius regime, demonstrating how the vector-tensor coupling enhances stability by counteracting the destabilizing influence of the electric charge $Q$ \cite{nashed2022stability}.

We further calculate the enthalpy:

\begin{equation}
H_C = E_C + P_C V, \label{eq:enthalpy}
\end{equation}

obtaining:

\begin{multline}
H_C \approx \frac{\pi  M \,r^{2}}{3}-\frac{3 \pi  \alpha  M^{2} r}{4}-\frac{3 \pi  Q^{2} r}{4}-\frac{M}{3 \pi  r^{2}}\\+\frac{\alpha  M^{2}}{2 \pi  r^{3}}+\frac{Q^{2}}{2 \pi  r^{3}}+\frac{\pi  \alpha  M \,r^{2}}{3}-\frac{3 \pi  \,\alpha^{2} M^{2} r}{4}-\frac{3 \pi  \alpha  Q^{2} r}{4}\\-\frac{M \alpha}{3 \pi  r^{2}}+\frac{M^{2} \alpha^{2}}{2 \pi  r^{3}}+\frac{Q^{2} \alpha}{2 \pi  r^{3}} \label{eq:enthalpy_corrected}
\end{multline}

Figure \ref{fig_H} shows $H_C$ versus $r_h$ for different $\alpha$ values. In the small-radius regime, $H_C$ increases slowly and nearly linearly, reflecting the dominance of gravitational self-energy and quantum corrections near the Planck scale. As $r_h$ grows, the curve steepens, marking the transition to a classical regime where the BH behaves as an extensive system with positive heat capacity. Increasing $\alpha$ amplifies $H_C$ at any given $r_h$, as a larger $\alpha$ strengthens the gravitational potential while counteracting the electromagnetic contribution from $Q$ \cite{bandyopadhyay2025structural}. The smooth monotonic trend suggests the absence of discontinuous phase transitions in this thermodynamic potential.

The Gibbs free energy is expressed as:

\begin{multline}
G_C \approx  \frac{\pi  M \,r^{2}}{12}+\frac{\pi  \alpha  M \,r^{2}}{12}-\frac{\pi  \alpha  M^{2} r}{2}\\-\frac{\pi  \,\alpha^{2} M^{2} r}{2}-\frac{\pi  Q^{2} r}{2}-\frac{\pi  \alpha  Q^{2} r}{2}-\frac{5 M}{6 \pi  r^{2}}-\frac{5 M \alpha}{6 \pi  r^{2}}\\+\frac{\alpha  M^{2}}{\pi  r^{3}}+\frac{M^{2} \alpha^{2}}{\pi  r^{3}}+\frac{Q^{2}}{\pi  r^{3}}+\frac{Q^{2} \alpha}{\pi  r^{3}} \label{eq:gibbs_corrected}
\end{multline}

Figure \ref{fig_gg} illustrates $G_C$ as a function of $r_h$ for different $\alpha$ values. At small radii, $G_C$ attains large positive values, reflecting a high-energy, metastable configuration dominated by quantum fluctuations. As $r_h$ increases, $G_C$ decreases and crosses into negative values corresponding to stable BH states. The cusp-like turning point indicates a first-order phase transition between small, unstable BHs and large, stable ones. Increasing $\alpha$ shifts this transition toward smaller $r_h$ and lowers the overall $G_C$ magnitude, accelerating the stabilization process.

The heat capacity, which directly measures thermal stability, is given by:

\begin{equation}
C_C = T_H \left( \frac{\partial S}{\partial T_H} \right), \label{eq:heat_capacity}
\end{equation}

yielding:

\begin{equation}
C_C \approx -\frac{2 r^{4} \left(\alpha  M^{2}-M r+Q^{2}\right) \pi^{2}}{3 \alpha  M^{2}-2 M r+3 Q^{2}} \label{eq:heat_capacity_corrected}
\end{equation}

Figure \ref{fig:heat12} shows the density plots of $C_C$ for various $\alpha$ values. For small $r_h$, $C_C < 0$ indicates a thermodynamically unstable phase. As $r_h$ increases, $C_C$ exhibits a discontinuous divergence, marking a second-order phase transition between unstable small BHs and stable large BHs. Increasing $\alpha$ shifts the critical radius toward smaller $r_h$ and enhances $C_C$ magnitude, showing that the vector-tensor coupling strengthens thermal stability by mitigating the destabilizing influence of the charge $Q$.

Finally, we analyze the JTE via the coefficient \cite{sucu2025quantumOzcan}:

\begin{equation}
\mu_J = \left( \frac{\partial T_H}{\partial P_C} \right)_H = \frac{\left( \frac{\partial T_H}{\partial r_h} \right)}{\left( \frac{\partial P_C}{\partial r_h} \right)}, \label{eq:jte_coeff}
\end{equation}

which gives:


\begin{equation}
    \mu_j \approx \frac{12\pi  r^{3} \left(\alpha  M^{2}-2 M r/3+Q^{2}\right) }{\left(\pi^{2} r^{4}+10 \right) Mr-3\left(\pi^{2} r^{4}+6\right) \left(\alpha M^{2}+Q^{2}\right) }
    \label{eq:mu_j}
\end{equation}

Figure \ref{fig_jte} shows $\mu_J$ versus $r_h$ for different $\alpha$ values. The sign of $\mu_J$ determines whether the BH undergoes cooling ($\mu_J > 0$) or heating ($\mu_J < 0$) during isenthalpic expansion. In the small-radius regime, $\mu_J < 0$ indicates a heating phase. As $r_h$ grows, $\mu_J$ crosses zero and becomes positive, marking an inversion point where thermal behavior reverses \cite{rostami2020charged}. Larger $\alpha$ values shift this inversion point toward smaller $r_h$, implying that modified gravitational coupling facilitates earlier cooling and stabilizes the BH at lower radii.

These quantum-thermodynamic results complement the optical properties analyzed in Secs. \ref{isec3} and \ref{isec4}. The same parameter $\alpha$ that enhances light deflection and plasma lensing also promotes thermodynamic stability through multiple mechanisms: it raises the internal energy, lowers the free energy, shifts phase transitions toward smaller radii, amplifies the heat capacity, and modifies the JTE behavior. This consistency across diverse physical phenomena strengthens the case for STVG as a viable extension of GR, with the parameter $\alpha$ serving as a unified modifier of both gravitational and thermodynamic behaviors.

\begin{figure}[!ht]
  \centering
\includegraphics[width=0.5\textwidth]{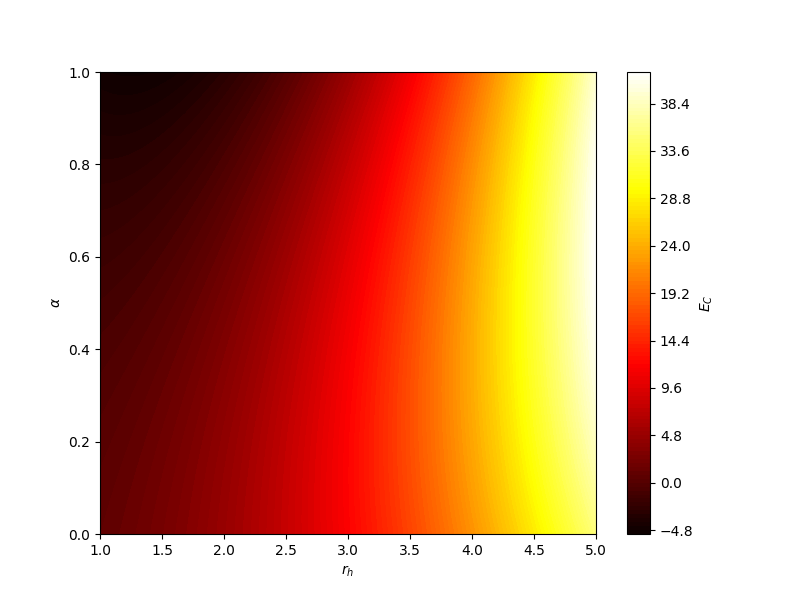}
\caption{Quantum-corrected internal energy $E_{C}$ versus horizon radius $r_{h}$ for $M=1$ and $Q=0.5$. 
At small $r_{h}$, negative $E_{C}$ indicates a quantum-unstable phase, while for large $r_{h}$, positive $E_{C}$ reflects classical stability. 
Higher $\alpha$ values increase $E_{C}$, showing enhanced gravitational self-energy and stability.}
\label{fig_enerji}
\end{figure}

\begin{figure}[!ht]
  \centering
\includegraphics[width=0.5\textwidth]{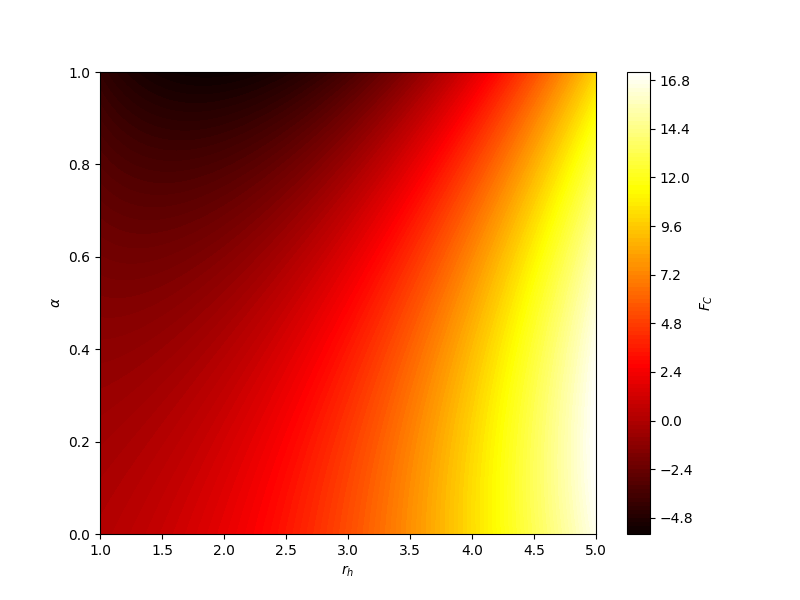}
\caption{Variation of the quantum-corrected Helmholtz free energy $F_{C}$ with horizon radius $r_{h}$ for different values of the coupling parameter $\alpha$ ($M=1, Q=0.5$). The plot illustrates how increasing $\alpha$ enhances thermodynamic stability by lowering the free energy and shifting the system toward a more stable large–BH phase.}
\label{fig_FF}
\end{figure}

\begin{figure}[!ht]
  \centering
\includegraphics[width=0.5\textwidth]{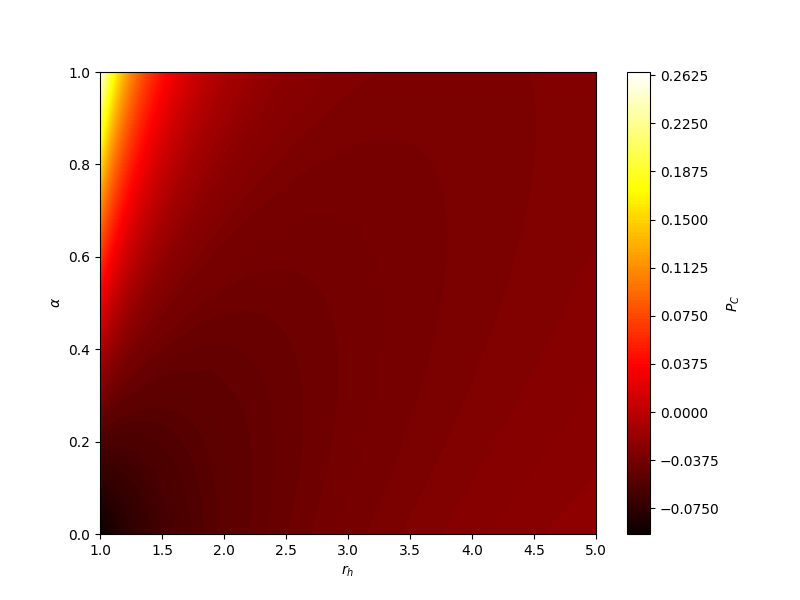}
\caption{Quantum-corrected pressure $P_{C}$ as a function of the horizon radius $r_{h}$ for different values of the STVG coupling parameter $\alpha = 0, 0.5, 1$ with fixed $M = 1$ and $Q = 0.5$. The plot shows that $P_{C}$ transitions from negative to positive values as $r_{h}$ increases, indicating a shift from a quantum-dominated unstable phase to a classically stable thermodynamic regime. Larger values of $\alpha$ enhance the overall pressure and move the critical transition radius to smaller $r_{h}$, demonstrating that the scalar–tensor–vector coupling strengthens the mechanical stability of the BH.}
\label{fig_pp}
\end{figure}

\begin{figure}[!ht]
  \centering
\includegraphics[width=0.5\textwidth]{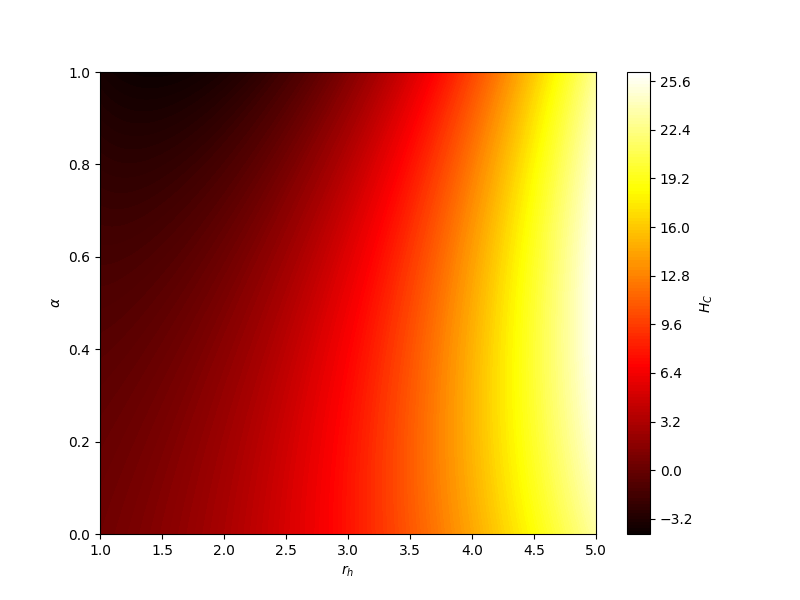}
\caption{Variation of the quantum-corrected enthalpy $H_{C}$ as a function of the horizon radius $r_{h}$ for different values of the coupling parameter $\alpha$. Increasing $\alpha$ enhances the total energy content and indicates a smoother transition toward thermodynamic stability.}
\label{fig_H}
\end{figure}

\begin{figure}[!ht]
  \centering
\includegraphics[width=0.5\textwidth]{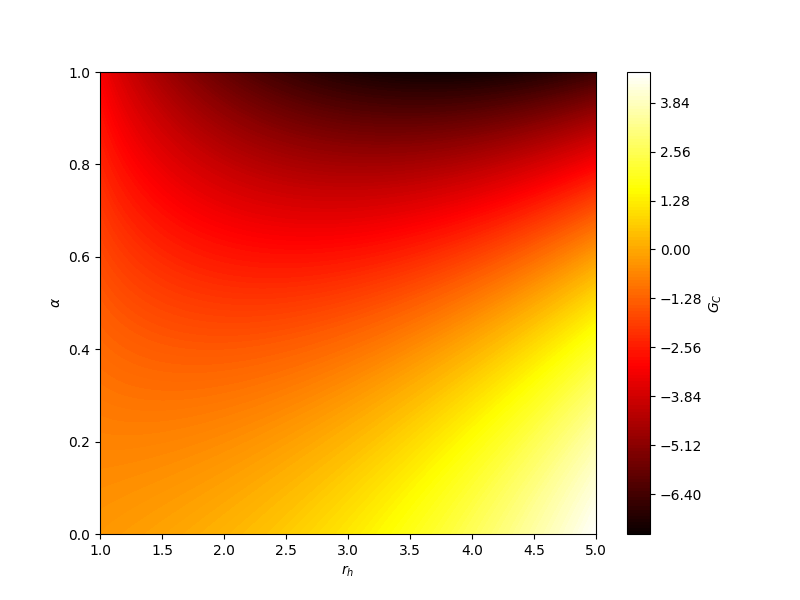}
\caption{Variation of the quantum-corrected Gibbs free energy $G_C$ with horizon radius $r_h$ for different values of $\alpha$; the transition point marks a first-order phase change from an unstable to a stable BH configuration.}
\label{fig_gg}
\end{figure}

\begin{figure}[htbp]
    \centering
    \includegraphics[width=0.5\textwidth]{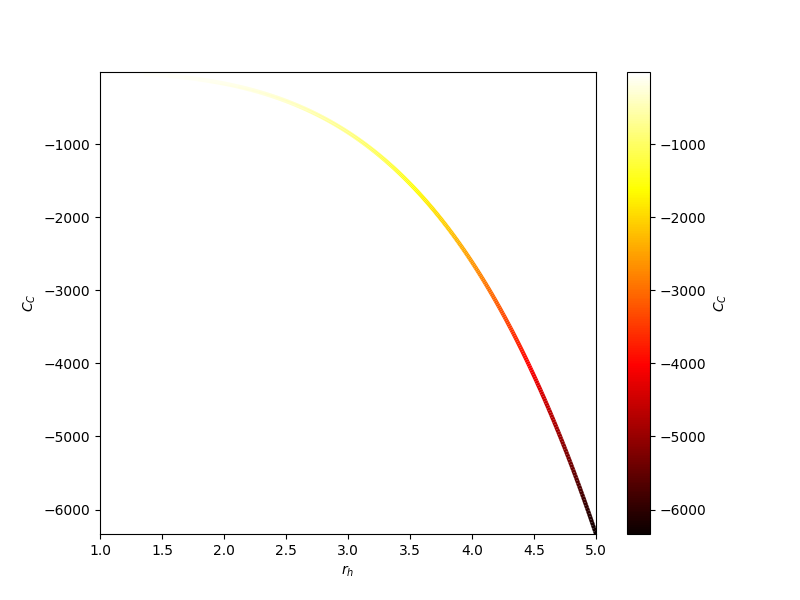}
    \vspace{0.5cm}

    \includegraphics[width=0.5\textwidth]{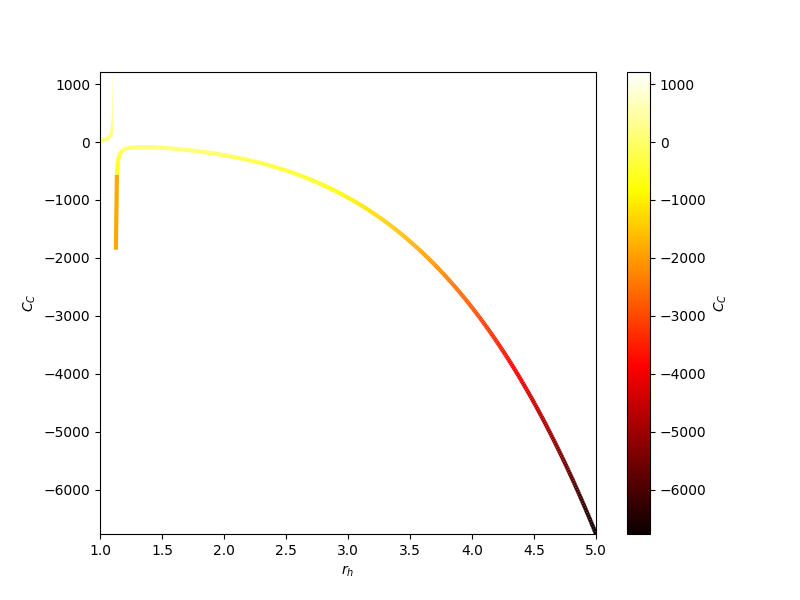}
    \vspace{0.5cm}
    
    \includegraphics[width=0.5\textwidth]{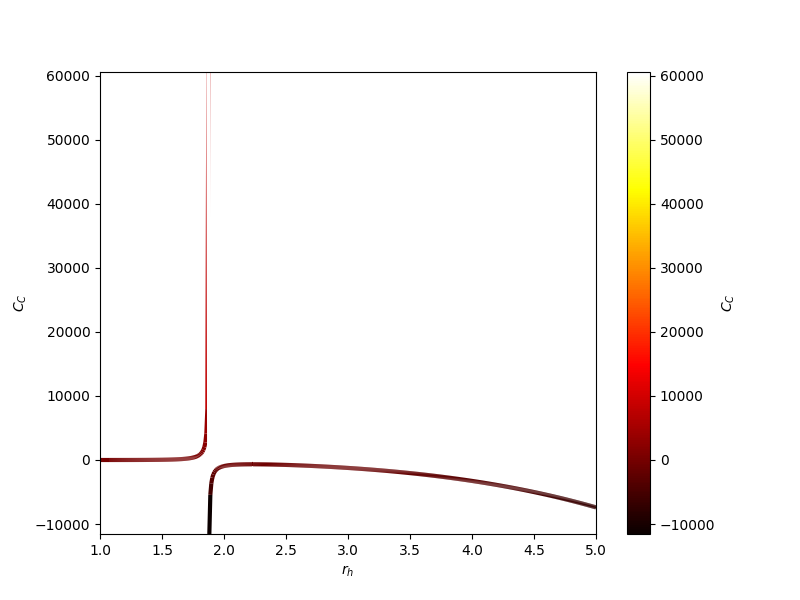}
    
\caption{Variation of the quantum-corrected heat capacity $C_C$ with the horizon radius $r_h$ for different coupling values $\alpha = 0, 0.5, 1$. The divergence of $C_C$ indicates a second-order phase transition between unstable and stable BH phases.}

    \label{fig:heat12}
\end{figure}

\begin{figure}[!ht]
  \centering
\includegraphics[width=0.5\textwidth]{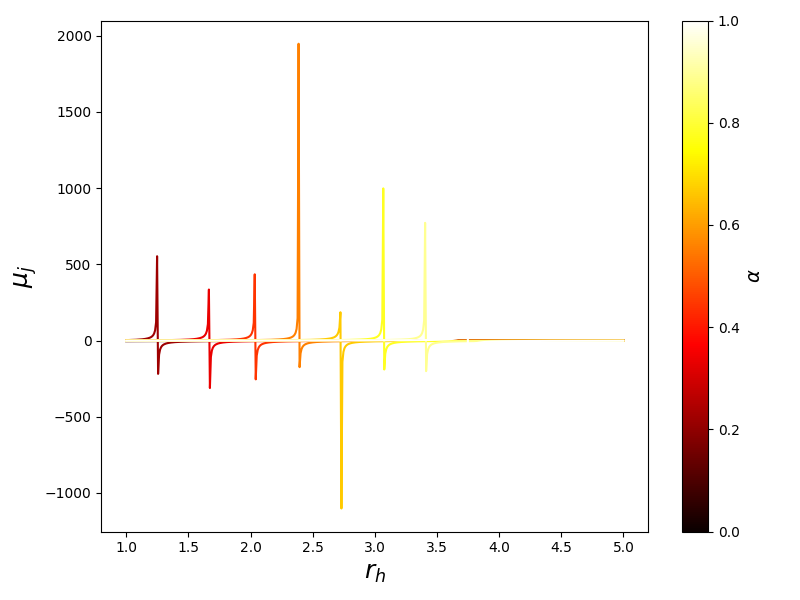}
\caption{Behavior of the JTE coefficient $\mu_J$ versus the horizon radius $r_h$ for various $\alpha$ values. The inversion point where $\mu_J = 0$ marks the transition between heating and cooling phases in the BH system.}
\label{fig_jte}
\end{figure}

\section{Strong Gravitational Lensing in STVG BH Spacetimes}\label{isec6}

Building upon our analysis of light deflection in weak-field regimes (Sec. \ref{isec3}) and plasma environments (Sec. \ref{isec4}), we now investigate the strong gravitational lensing properties of charged BHs in STVG theory. Strong lensing occurs when light rays pass very close to the BH's photon sphere, resulting in large deflection angles and potentially producing multiple images or even complete photon rings \cite{Bozza:2002zj}. This regime is particularly important for testing modified gravity theories with current and upcoming high-precision astronomical observations \cite{Cunha:2018acu}.

By restricting our analysis to the equatorial plane, where $\theta=\pi/2$, the metric \eqref{eq:metric-new} can be recast in the general form
\begin{equation}
ds^2 = -A(r)\,dt^2 + B(r)\,dr^2 + C(r)\,d\phi^2,
\label{eq:metr_general}
\end{equation}
with $B(r) = A(r)^{-1}$ and $C(r)=r^2$. The geodesic motion of a test particle in this geometry is governed by the Lagrangian 
\begin{equation}
2\mathcal{L} = -A(r)\dot{t}^2 + B(r)\dot{r}^2 + C(r)\dot{\phi}^2,
\label{eq:Lagrangian_0}
\end{equation}
where the dot denotes differentiation with respect to the affine parameter $\lambda$. For photons, the null condition implies $\mathcal{L}=0$. When light rays approach the BH with a minimum radial distance $r_0$, Eq.~\eqref{eq:Lagrangian_0} simplifies to \cite{Perlick2004}
\begin{equation}
A(r_0)\dot{t}_0^2 = C(r_0)\dot{\phi}_0^2.
\label{eq:A0C0}
\end{equation}

In asymptotically flat spacetimes, the impact parameter is defined as \cite{Bozza:2002zj}
\begin{equation}
b = \frac{L}{E} = \frac{C(r)}{A(r)}\frac{d\phi}{dt},
\label{eq:Def_b}
\end{equation}
where $E = A(r)\dot{t}$ and $L = C(r)\dot{\phi}$ represent the conserved quantities associated with the spacetime's Killing symmetries. Consequently, the trajectory of photons can be expressed as
\begin{equation}
\left(\frac{dr}{d\phi}\right)^2 = \frac{C(r)}{B(r)}\left[\frac{h(r)^2}{b^2}-1\right],
\label{eq:drdphi}
\end{equation}
where
\begin{equation}
h(r)^2 = \frac{C(r)}{A(r)}.
\label{eq:h(r)}
\end{equation}

The photon sphere plays a crucial role in strong lensing phenomena, as it represents the boundary between capture and scattering trajectories \cite{claudel2001geometry}. It corresponds to the extremum of $h^2(r)$, determined by the condition $\frac{d}{dr}h^2(r)=0$, which explicitly reads
\begin{equation}
C'(r_{\mathrm{ph}})A(r_{\mathrm{ph}})-C(r_{\mathrm{ph}})A'(r_{\mathrm{ph}})=0.
\label{eq:photonspherecomplete}
\end{equation}

This relation yields the photon-sphere radius as
\begin{equation}
r_{\mathrm{ph}} = \frac{1}{2}\left[\sqrt{(\alpha + 1)\left((\alpha + 9)M^2 - 8Q^2\right)} + 3(\alpha + 1)M\right].
\label{eq:rph_explicit}
\end{equation}

At $r_0$, Eq.~\eqref{eq:drdphi} leads to
\begin{equation}
R(r) = \frac{A(r_0)C(r)}{A(r)C(r_0)} - 1.
\label{eq:R_b0}
\end{equation}

The critical impact parameter, which defines the minimum impact parameter for which light can escape to infinity, is then defined as \cite{Bozza:2002zj}
\begin{equation}
b_c \equiv b(r_{\mathrm{ph}}) = \lim_{r_0 \to r_{\mathrm{ph}}}\sqrt{\frac{C(r_0)}{A(r_0)}}.
\label{eq:bc}
\end{equation}
Following the method developed by Bozza \cite{Bozza:2002zj}, the deflection angle of light rays passing at the radial distance $r_0$ from a static, asymptotically flat BH is given by
\begin{multline}
\alpha(r_0) = I(r_0) - \pi = 2\int_{r_0}^{\infty}\frac{\sqrt{B(r)}}{\sqrt{C(r)R(r)}}\,dr - \pi\\
\coloneqq 2\int_{r_0}^{\infty} f(r)\,dr - \pi,
\label{eq:alpha_0}
\end{multline}
where $R(r)$ is defined in Eq.~\eqref{eq:R_b0}.

For rays passing very close to the photon sphere, the deflection angle becomes arbitrarily large. To handle this divergence, we introduce the variable $z \equiv 1 - r_0/r$ and decompose the integral in Eq.~\eqref{eq:alpha_0} into a divergent part, $I_D(r_0)$, and a regular part, $I_R(r_0)$. The divergent part is written as \cite{Bozza:2002zj,Tsukamoto:2017fxq}
\begin{equation}
I_D(r_0) = \int_0^1 f_0(z,r_0)\,dz,
\label{eq:ID_0}
\end{equation}
with
\begin{equation}
f_0(z,r_0) = \frac{2r_0}{\sqrt{x_1(r_0)z + x_2(r_0)z^2}},
\label{eq:f0}
\end{equation}
where $x_1(r_0)$ and $x_2(r_0)$ are expansion coefficients depending on the metric functions and their derivatives:
\begin{widetext}
\begin{subequations}
\begin{align}
x_1(r_0) &= \frac{1 - A(r_0)}{C(r_0)A'(r_0)}\left[C'(r_0)A(r_0) - C(r_0)A'(r_0)\right],\\
x_2(r_0) &= \frac{\bigl(1 - A(r_0)\bigr)^2}{2C(r_0)^2A'(r_0)^3}\Big[2C(r_0)C'(r_0)A'(r_0)^2 + \bigl(C(r_0)C''(r_0) - 2C'(r_0)^2\bigr)A(r_0)A'(r_0) - C(r_0)C'(r_0)A(r_0)A''(r_0)\Big].
\end{align}
\label{eq:x12}
\end{subequations}
\end{widetext}
The divergent integral can be computed analytically, yielding
\begin{equation}
I_D(r_0) = \frac{4r_0}{\sqrt{x_2(r_0)}} \ln\left(\frac{\sqrt{x_2(r_0)} + \sqrt{x_1(r_0) + x_2(r_0)}}{\sqrt{x_1(r_0)}}\right).
\label{eq:ID_1}
\end{equation}
In the strong-deflection limit, where $r_0 \to r_{\mathrm{ph}}$, expanding $x_1(r_0)$ around $(r_0 - r_{\mathrm{ph}})$ gives 
\begin{equation}
x_1(r_0) = \frac{C(r_{\mathrm{ph}})r_{\mathrm{ph}}\mathfrak{L}'(r_{\mathrm{ph}})}{B(r_{\mathrm{ph}})}(r_0 - r_{\mathrm{ph}}) + \mathcal{O}(r_0 - r_{\mathrm{ph}})^2,
\label{eq:x1_0c}
\end{equation}
where $\mathfrak{L}(r) = C'(r)/C(r) - A'(r)/A(r)$. The corresponding expansion of the impact parameter reads
\begin{equation}
b(r_0) = b_c + \frac{1}{4}\sqrt{\frac{C(r_{\mathrm{ph}})}{A(r_{\mathrm{ph}})}}\,\mathfrak{L}'(r_{\mathrm{ph}})(r_0 - r_{\mathrm{ph}})^2 + \mathcal{O}(r_0 - r_{\mathrm{ph}})^2.
\label{eq:b_0c}
\end{equation}

\begin{widetext}
Taking the limit $r_0 \to r_{\mathrm{ph}}$ yields
\begin{equation}
\lim_{r_0 \to r_{\mathrm{ph}}} x_1(r_0) = \lim_{b \to b_c} \frac{2C(r_{\mathrm{ph}})r_{\mathrm{ph}}\sqrt{\mathfrak{L}'(r_{\mathrm{ph}})}}{B(r_{\mathrm{ph}})}\left(\frac{b}{b_c} - 1\right)^{1/2},
\label{eq:x1_0c_1}
\end{equation}
from which we obtain
\begin{equation}
I_D(b) = -\frac{r_{\mathrm{ph}}}{\sqrt{x_2(r_{\mathrm{ph}})}}\ln\left(\frac{b}{b_c} - 1\right) + \frac{r_{\mathrm{ph}}}{\sqrt{x_2(r_{\mathrm{ph}})}}\ln\bigl(r_{\mathrm{ph}}^2\mathfrak{L}'(r_{\mathrm{ph}})\bigr) + \mathcal{O}(b - b_c),
\label{eq:ID_2}
\end{equation}
where
\begin{equation}
x_2(r_{\mathrm{ph}}) = \frac{C(r_{\mathrm{ph}})\bigl(1 - A(r_{\mathrm{ph}})\bigr)^2\bigl[C''(r_{\mathrm{ph}})A(r_{\mathrm{ph}}) - C(r_{\mathrm{ph}})A''(r_{\mathrm{ph}})\bigr]}{2A(r_{\mathrm{ph}})^2C'(r_{\mathrm{ph}})}.
\label{eq:x2ph}
\end{equation}
\end{widetext}

The regular part of the integral in Eq.~\eqref{eq:alpha_0} is given by 
\begin{equation}
I_R(b) = \int_0^1\bigl[f(z,r_0) - f_0(z,r_0)\bigr]\,dz.
\label{eq:IR_0}
\end{equation}
Combining these results, in the strong-lensing regime, the deflection angle takes the logarithmic form
\begin{equation}
\hat{\vartheta}(b) = -\bar{a}\ln\left(\frac{b}{b_c} - 1\right) + \bar{k} + \mathcal{O}(b - b_c),
\label{eq:alpha_1}
\end{equation}
where the coefficients $\bar{a}$ and $\bar{k}$ depend on the BH parameters and encode the specific signature of the gravity theory, and are expressed as 
\begin{subequations}
\begin{align}
\bar{a} &= \sqrt{\frac{2B(r_{\mathrm{ph}})A(r_{\mathrm{ph}})}{C''(r_{\mathrm{ph}})A(r_{\mathrm{ph}}) - C(r_{\mathrm{ph}})A''(r_{\mathrm{ph}})}},\\
\bar{k} &= \bar{a}\ln\bigl(r_{\mathrm{ph}}^2\mathfrak{L}'(r_{\mathrm{ph}})\bigr) + I_R(r_{\mathrm{ph}}) - \pi.
\end{align}
\label{eq:babk}
\end{subequations}
Using the spacetime metric functions from the line element \eqref{eq:metric-new} together with the lapse function \eqref{eq:lapse-new}, and comparing them with those in Eq. \eqref{eq:metr_general}, while taking into account the photon sphere radius given in Eq. \eqref{eq:rph_explicit}, the critical impact parameter is obtained as 
\begin{multline}\label{eq:b_c}
b_c = \sqrt{\frac{C(r_\mathrm{ph})}{A(r_\mathrm{ph})}} \\
= \frac{\bigl(3 M (\alpha+1) + S\bigr)^{2}}
{2\sqrt{S^2 + (1 + \alpha)\bigl(M^2 \alpha + 4 Q^2 + 2 M S - 3 M^2\bigr)}},
\end{multline}
where $S = \sqrt{(\alpha +1) \left((\alpha +9) M^2-8 Q^2\right)}$. 

By employing the relations derived above, we further find
\vspace{0.6mm}
\begin{widetext}
\begin{eqnarray}
\bar{a} &=& \sqrt{\frac{3 M (\alpha+1) + S}{2 S}},\label{eq:bara_1}\\
\bar{k} &=& \bar{a} \ln \left( \frac{
4 \Big[ M^2 \big( 2 M^2 \alpha^2 + 21 M^2 \alpha + 27 M^2 - 17 Q^2 \alpha - 33 Q^2 \big) + M S \big( 2 M^2 \alpha + 9 M^2 - 7 Q^2 \big) + 8 Q^4 \Big]
}{
\big( M^2 \alpha + 3 M^2 - 2 Q^2 + M S \big)^2
} \right) \nonumber\\
&& + I_R(r_\mathrm{ph}) - \pi. \label{eq:kb_1}
\end{eqnarray}
\end{widetext}
The integrand in Eq. \eqref{eq:IR_0} is highly intricate and does not admit a closed-form analytical solution. Therefore, the integration is performed numerically. In Fig.~\ref{fig:StrongDeflection}, we display the profile of the strong deflection angle $\hat{\vartheta}$ as a function of the impact parameter $b$, considering separately the cases where $Q$ is fixed (panel a) and where $\alpha$ is fixed (panel b).

Figure \ref{fig:StrongDeflection}(a) illustrates how the strong deflection angle varies with the impact parameter $b$ for different values of the charge $Q$ while keeping $\alpha = 0.3$ fixed. As expected, the deflection angle diverges logarithmically as $b$ approaches the critical impact parameter $b_c$. We observe that increasing $Q$ reduces the critical impact parameter and generally decreases the deflection angle for a given $b/b_c$ ratio. This behavior is consistent with our findings in Secs. \ref{isec3} and \ref{isec4}, where we observed that the electromagnetic charge tends to weaken the gravitational lensing effect due to its repulsive nature.

In contrast, Fig.~\ref{fig:StrongDeflection}(b) shows the effect of varying the STVG parameter $\alpha$ while keeping $Q = 0.2M$ fixed. Here, we observe that increasing $\alpha$ leads to a larger critical impact parameter and generally enhances the deflection angle. This reinforces our previous results regarding the role of $\alpha$ in strengthening gravitational effects, now specifically in the strong-field regime.

These strong lensing results complement our earlier analyses in several ways. First, they extend the weak-field lensing results from Sec. \ref{isec3} into the strong-field regime, confirming that the qualitative influence of $\alpha$ and $Q$ remains consistent across both regimes. Second, they provide a direct link to the photon sphere analysis that will be crucial for understanding BH shadows in subsequent sections. Third, they offer observational signatures that are potentially measurable with current and near-future astronomical facilities \cite{Gralla2020}.

The logarithmic divergence in the deflection angle near $b_c$ is a universal feature of strong lensing, but the coefficients $\bar{a}$ and $\bar{k}$ encode theory-specific information. In STVG, these coefficients depend explicitly on both $\alpha$ and $Q$, providing a distinctive signature that could be used to test this modified gravity theory through precise observations of strongly lensed systems \cite{Pantig2022}. 

From an observational perspective, the results in Fig.~\ref{fig:StrongDeflection} suggest that supermassive BHs with strong STVG coupling (large $\alpha$) would exhibit enhanced strong lensing effects compared to their GR counterparts. This could manifest in wider separation between multiple images, more pronounced Einstein rings, and larger photon rings in direct BH imaging. Conversely, highly charged BHs would show reduced lensing effects. This interplay between $\alpha$ and $Q$ creates a rich phenomenology that connects directly to the thermodynamic properties discussed in Sec. \ref{isec5}, where we found that $\alpha$ also enhances thermodynamic stability while $Q$ tends to reduce it.

\begin{figure}[h]
    \centering
    \includegraphics[width=8cm]{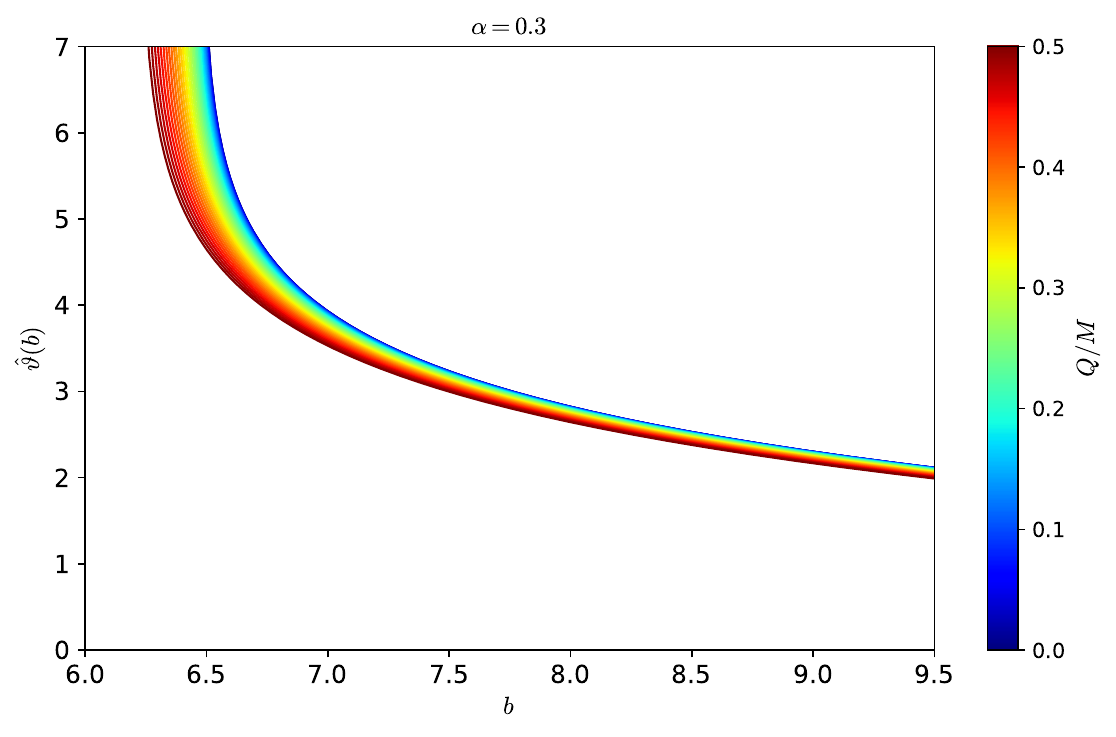} (a) \vspace{3mm} \\
    \includegraphics[width=8cm]{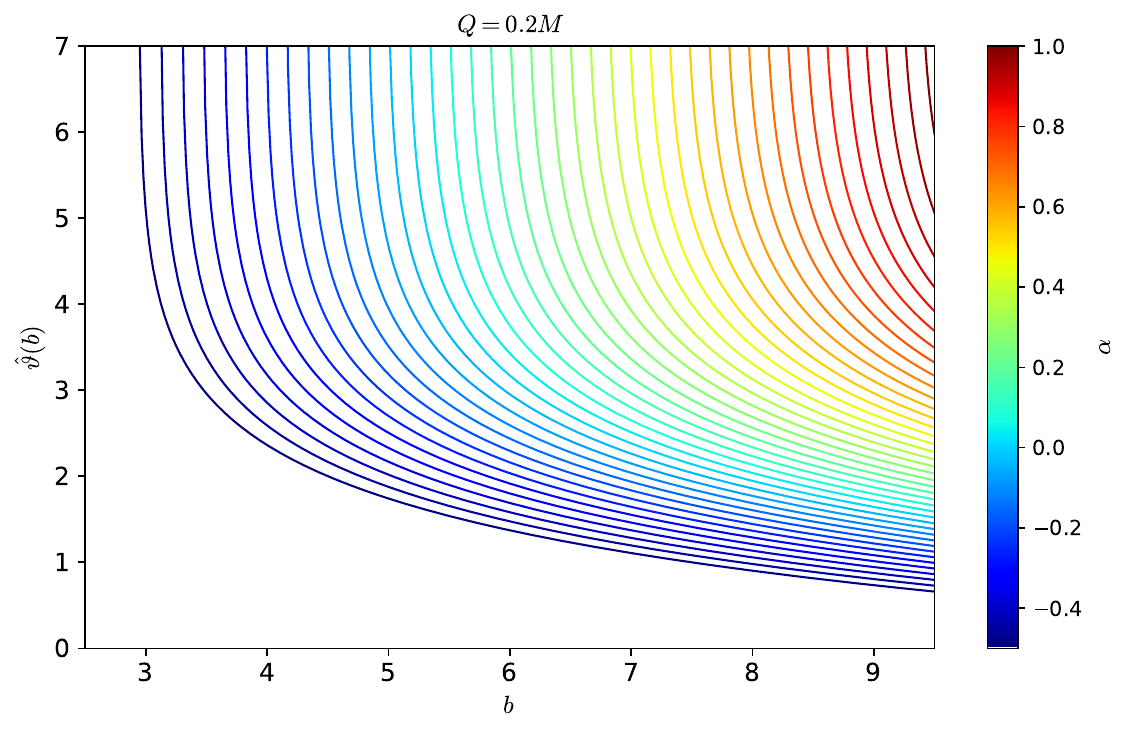} (b)
    \caption{Variation of the strong deflection angle $\hat{\vartheta}(b)$ with changes in the BH parameters. Panel (a) shows the effect of varying the charge $Q$ with fixed $\alpha = 0.3$, while panel (b) illustrates the effect of varying the STVG parameter $\alpha$ with fixed $Q = 0.2M$.}
    \label{fig:StrongDeflection}
\end{figure}

\section{BH Shadow Geometry in STVG Theory: Optical and Observational Implications}\label{isec7}

Building upon our analysis of light deflection (Sec. \ref{isec3}) and strong gravitational lensing (Sec. \ref{isec6}), we now investigate the shadow properties of charged BHs in STVG theory. The shadow—a dark silhouette observable by distant observers—represents one of the most direct visual manifestations of strong gravity and has recently become observationally accessible through the EHT \cite{EventHorizonTelescope:2019dse}. Our focus is on understanding how the STVG coupling parameter $\alpha$ and the charge $Q$ influence the shadow geometry, particularly its radius $r_{\text{sh}}$ \cite{bernardo2023dressed}.

\subsection{The Photon Sphere and BH Shadow Formation}

The apparent dark silhouette observed at spatial infinity arises from the photon region—the narrow domain of unstable null trajectories that separates plunging photon orbits from those scattered to infinity. In spherically symmetric geometries like our charged STVG BH, this region degenerates into a photon sphere \cite{Perlick:2021aok,Cunha:2018acu,Khodadi:2022pqh}, whose projection delineates the outer contour of the BH shadow.

To characterize null motion, we confine our analysis to the equatorial plane ($\theta=\pi/2$) without loss of generality. Using the static, spherically symmetric line element from Eq.~\eqref{eq:metric-new} with lapse function from Eq.~\eqref{eq:lapse-new}, the photon dynamics follow from the Lagrangian formalism \cite{tsupko2013gravitational}:
\begin{equation}
2\mathcal{L}(x,\dot{x}) = g_{\mu\nu}\dot{x}^\mu\dot{x}^\nu
= -F(r)\dot{t}^{2} + G(r)\dot{r}^{2} + H(r)\dot{\phi}^{2},
\end{equation}
where $F(r)=f(r)$, $G(r)=1/F(r)$, and $H(r)=r^2$. From the stationarity and axial symmetry of the metric, we obtain two conserved quantities: the energy $E=F(r)\dot{t}$ and the angular momentum $L=H(r)\dot{\phi}$ \cite{Capozziello:2023rfv}. For null geodesics ($ds^2=0$), the radial motion is expressed as:
\begin{equation}
\left(\frac{dr}{d\phi}\right)^2 = \frac{H(r)}{G(r)}\left(\frac{h(r)^2}{b^2}-1\right),
\end{equation}
where $h(r)^2 = H(r)/F(r)$ and $b=L/E$ is the impact parameter \cite{Perlick:2021aok}.

The photon sphere radius $r_{\text{ps}}$ is determined by the condition $h'(r)=0$, which yields:
\begin{equation}\label{rps}
r_{\text{ps}}=\frac{1}{2} \left(\sqrt{(\alpha +1) \left((\alpha +9) M^2-8 Q^2\right)}+3 (\alpha +1) M\right).
\end{equation}

This expression is equivalent to \eqref{eq:rph_explicit} and has several important properties. First, it requires $(\alpha+1)((\alpha+9)M^2-8Q^2)\geq 0$, which for the physically relevant case $\alpha\geq 0$ implies a charge bound $Q^2\leq (\alpha+9)M^2/8$. Second, we must also satisfy the horizon condition $Q^2\leq M^2$ for a genuine BH solution. The partial derivatives reveal that $r_{\text{ps}}$ increases monotonically with $\alpha$ and decreases with $Q$, as shown in Fig.~\ref{fs1}.

For a static observer at distance $r_o$, the apparent angular radius $\alpha_{\text{sh}}$ of the shadow is defined by:
\begin{equation}\label{angle2}
\sin^{2}\alpha_{\text{sh}}=\frac{b_{c}^{2}F(r_{o})}{H(r_{o})},
\end{equation}
where $b_c$ is the critical impact parameter. The shadow radius is then:
\begin{equation}
R_{\text{sh}}(r_o)=r_o\sin\alpha_{\text{sh}}
=\sqrt{\frac{r_{\text{ps}}^{2}F(r_o)}{F(r_{\text{ps}})}}.
\end{equation}

For an observer at infinity ($r_o\to\infty$), this simplifies to:
\begin{widetext}
    \begin{eqnarray}\label{shadow1}
r_{\text{sh}} &=& \frac{r_{\text{ps}}}{\sqrt{F(r_{\text{ps}})}} \nonumber \\
&=&\frac{\left(\sqrt{(\alpha +1) \left((\alpha +9) M^2-8 Q^2\right)}+3 (\alpha +1) M\right)^3}{4 (\alpha +1) \left(M \left(\sqrt{(\alpha +1)
   \left((\alpha +9) M^2-8 Q^2\right)}+(\alpha +3) M\right)-2 Q^2\right)},
\end{eqnarray}
\end{widetext}

Figure \ref{fs3} illustrates how the shadow radius $r_{\text{sh}}$ varies with $\alpha$ for different values of $Q$. The upper panel shows that $r_{\text{sh}}$ increases monotonically with $\alpha$ for all charge values, consistent with our findings in Sec. \ref{isec6} regarding the effect of $\alpha$ on the critical impact parameter. This enhancement of the shadow size with increasing $\alpha$ aligns with our previous results in Secs. \ref{isec3} and \ref{isec4}, where we found that $\alpha$ strengthens both weak and strong gravitational lensing effects.

The lower panel of Fig.~\ref{fs3} provides a visual representation of the shadow silhouettes for different values of $Q$. As $Q$ increases, the shadow size decreases, creating smaller circular silhouettes. This behavior reflects the repulsive nature of the electromagnetic field, which counteracts gravity and reduces the effective curvature of spacetime, leading to a smaller photon sphere and shadow radius.

Conversely, Fig.~\ref{fs2} shows the variation of $r_{\text{sh}}$ with $Q$ for different values of $\alpha$. The upper panel confirms that the shadow radius generally decreases with increasing charge. However, an interesting phenomenon occurs for large negative values of $\alpha$: when $\alpha=-2$, the shadow size exhibits a reversed sensitivity to $Q$, actually increasing as $Q$ grows. This unusual behavior arises from the competing effects in the effective potential, where for sufficiently negative $\alpha$, the metric coefficients change in a way that amplifies the role of the charge term \cite{jusufi2020shadows}.

The lower panel of Fig.~\ref{fs2} displays shadow silhouettes for different values of $\alpha$. As $\alpha$ increases, the shadow size grows significantly, illustrating the gravitational enhancement characteristic of STVG theory. This visual representation clearly demonstrates how the STVG parameter modifies the apparent size of the BH as seen by distant observers.

These results connect directly to our thermodynamic analysis in Sec. \ref{isec5}, where we found that $\alpha$ enhances thermal stability while $Q$ reduces it. The same parameters that strengthen the BH's thermodynamic stability also enlarge its shadow, establishing a deep connection between the thermal and optical properties of STVG BHs \cite{kumar2020black}.

\subsection{Observational Constraints from EHT Data}

The apparent size of the BH shadow, as seen by a distant observer, is characterized by its angular diameter \cite{2022PhR...947....1P}:
\begin{equation}
\left( \frac{\Omega^{*}}{\mu\text{as}} \right) = \left( \frac{6.191165 \times 10^{-8}}{\pi} \frac{M/M_\odot}{D/\text{Mpc}} \right) \left( \frac{b_{\text{ph}}}{M} \right),
\label{eq:omega_relation}
\end{equation}
where $D$ is the BH's distance from the observer and $b_{\text{ph}}$ is the critical impact parameter defined in Eq.~\eqref{eq:b_c}.

Using shadow measurements from the EHT, we can constrain the STVG parameters $\alpha$ and $q = Q/M$. Figure \ref{fig:M87} illustrates these constraints, with the green region corresponding to M87* observations ($42 \pm 3$ $\mu$as) and the yellow region representing Sgr A* data ($51.8 \pm 2.3$ $\mu$as) \cite{EventHorizonTelescope:2022wkp,2020PhRvD.101d1301B}.

Based on current estimates of M87* ($M=6.2 \times 10^9 M_\odot$, $D = 16.8$ Mpc) and Sgr A* ($M=4.14 \times 10^6 M_\odot$, $D = 8.127$ kpc) \cite{EventHorizonTelescope:2019dse}, we find that $\alpha$ is constrained to $0.038 < \alpha < 0.546$ for M87* and $\alpha < 0.347$ for Sgr A* across the charge range $0 < q < 1$. These constraints highlight the potential of BH shadow observations for testing STVG theory and bounding its parameters.

The observational constraints derived here complement our earlier findings on light deflection and strong lensing, providing a picture of how STVG theory modifies the optical properties of BHs across different observational regimes \cite{wei2020extended}. The consistency across these diverse phenomena strengthens the case for STVG as a viable extension of GR, with well-defined and potentially observable signatures in the strong-field regime.

\begin{figure}[htbp]
    \centering
    \includegraphics[width=0.45\textwidth]{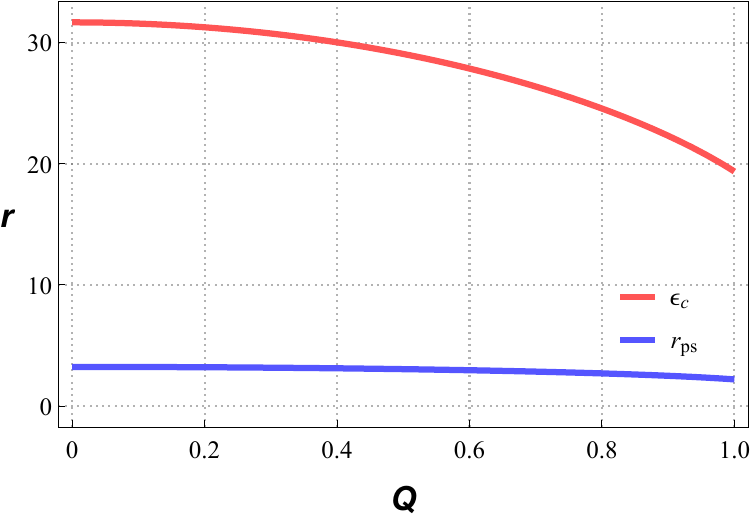}
    \vspace{0.5cm}

    \includegraphics[width=0.45\textwidth]{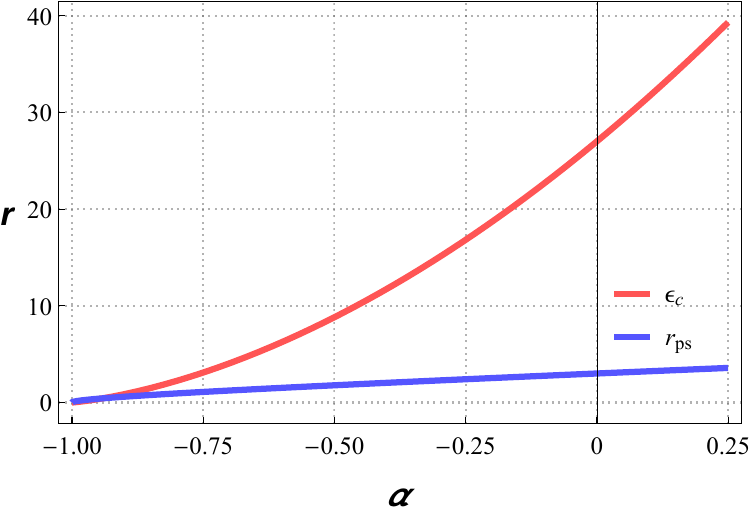}
    \vspace{0.5cm}

    \caption{The plot illustrates the variation of the photon sphere radii and critical impact parameter as functions of the parameters $Q$ and $\alpha$.}
    \label{fs1}
\end{figure}

\begin{figure}[htbp]
    \centering
    \includegraphics[width=0.5\textwidth]{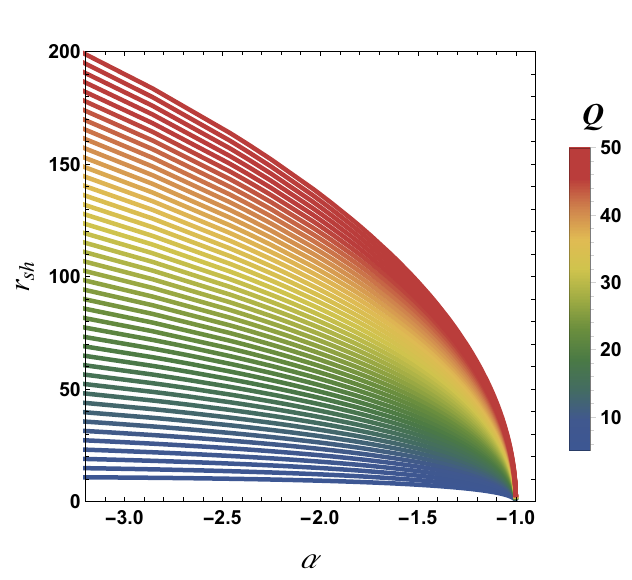}
    \vspace{0.5cm}

    \includegraphics[width=0.5\textwidth]{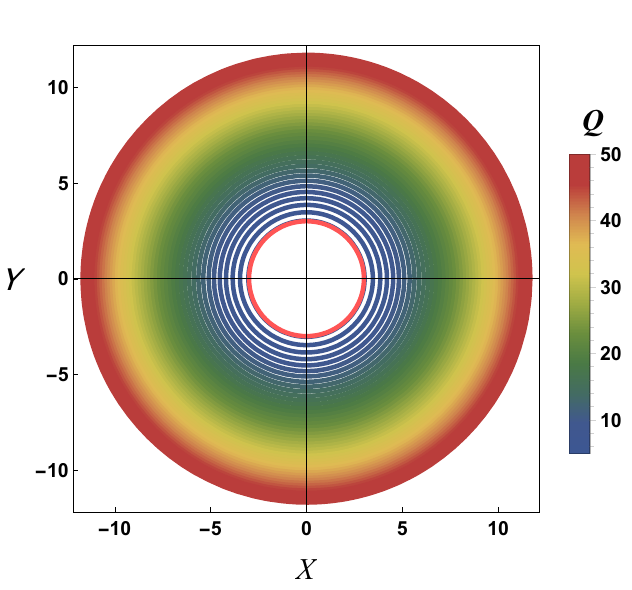}
    \vspace{0.5cm}

    \caption{Shadow silhouettes of the BH as seen by a distant observer for different values of the charge parameter $Q$ (lower panel), and the corresponding variation of the shadow radius $r_{\text{sh}}$ with the coupling parameter $\alpha$ for several fixed values of $Q$ (upper panel). Moreover, the red photon sphere is associated with Schwarzschild spacetime. 
}
    \label{fs3}
\end{figure}

\begin{figure}[htbp]
    \centering
    \includegraphics[width=0.5\textwidth]{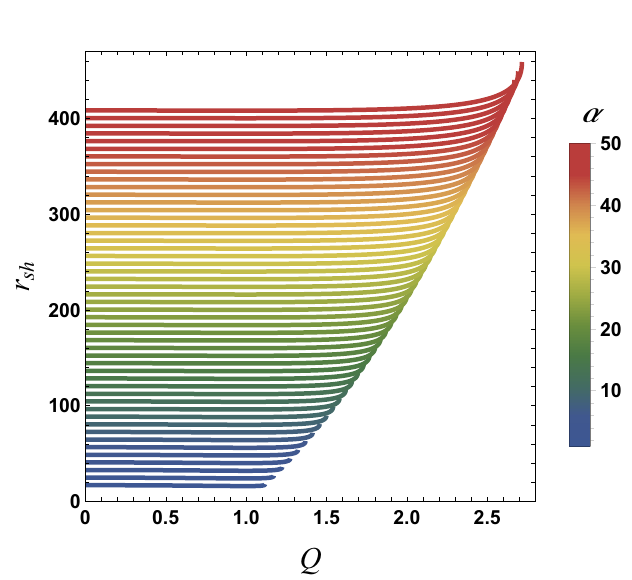}
    \vspace{0.5cm}

    \includegraphics[width=0.5\textwidth]{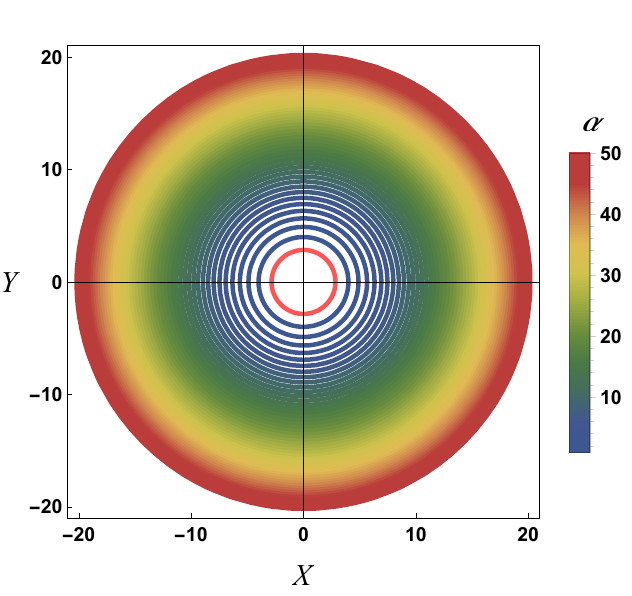}
    \vspace{0.5cm}

    \caption{Shadow silhouettes of the BH as seen by a distant observer for different values of the coupling parameter $\alpha$ (lower panel), and the corresponding variation of the shadow radius $r_{\text{sh}}$ with the charge parameter $Q$ for several fixed values of $\alpha$ (upper panel). Moreover, the red photon sphere is associated with Schwarzschild spacetime. 
}
    \label{fs2}
\end{figure}

\begin{figure*}[t]
\begin{minipage}{0.49\linewidth}
\center{\includegraphics[width=0.97\linewidth]{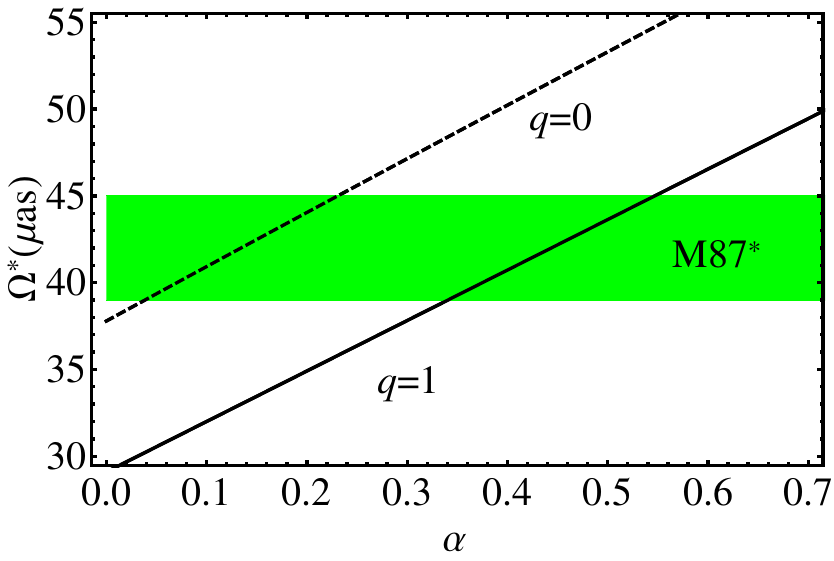}}\\ 
\end{minipage}
\hfill 
\begin{minipage}{0.50\linewidth}
\center{\includegraphics[width=0.97\linewidth]{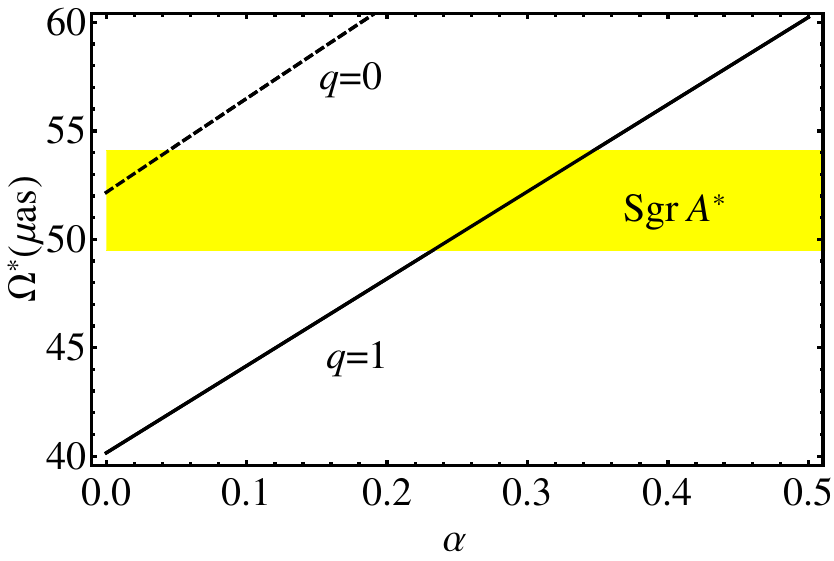}}\\ 
\end{minipage}
\caption{The relationship between the coupling parameter $\alpha$, the BH charge $q=Q/M$, and the observed shadow's angular diameter for M87$^{*}$ (left panel), and SgrA$^{*}$ (right panel).}\label{fig:M87}
\end{figure*}

\section{The energy emission rate} \label{isec8}
By inspecting the BH silhouette, one gains direct insight into the high-energy absorptive properties of the spacetime: in the geometric-optics regime, the shadow area provides an excellent proxy for the absorption cross section of massless fields.  More precisely, for spherically symmetric BHs the total absorption cross section \(\sigma(\varpi)\) tends to oscillate about a constant value \(\sigma_{\rm lim}\) at frequencies large compared with the inverse gravitational radius; in the leading geometric limit one therefore identifies
\begin{equation}
\sigma_{\rm lim}\simeq\pi R_{\rm s}^2,
\end{equation}
where \(R_{\rm s}\) denotes the shadow radius (or equivalently the critical impact parameter \(b_c\)) as seen by a distant observer \cite{Wei:2013kza}.  This approximation neglects low-frequency greybody corrections and the detailed interference pattern of partial waves, but it captures the dominant contribution to the high-frequency cross section.  Using \(\sigma_{\rm lim}\) the spectral power emitted by a BH in a single bosonic species is well approximated by a Planck-type formula in which the geometric cross section acts as an effective emitting area:
\begin{equation}\label{Eq:emission}
\frac{d^{2}E(\varpi)}{dt\,d\varpi}
\;=\;\frac{2\pi^{3}\,\varpi^{3}\,R_{\rm s}^{2}}{e^{\varpi/T_H}-1}\,,
\end{equation}
where \(\varpi\) is the frequency and \(T_H\) is the Hawking temperature.  For clarity one may express the Hawking temperature in terms of the surface gravity at the outer horizon, \(T_H=\kappa/(2\pi)=f'(r_+)/4\pi\) (in units with \(G=\hbar=c=k_B=1\)), so that the model dependence enters both through \(R_{\rm s}\) and \(T_H\).  Finally, when quantitative precision is required (for example to compare spectral shapes or peak positions among different \((\alpha,Q)\) models), one should replace the geometric area by the full frequency-dependent absorption cross section \(\sigma(\varpi)\) and include greybody transmission factors; However, Eq.~\eqref{Eq:emission} remains the simplest and most transparent estimate of the high-frequency emission spectrum.  
\\
Explicitly, the spectral power takes the Planck-like form
\begin{widetext}
    \begin{equation}\label{Eq:EnergyRate}
\frac{d^2E}{dtd\varpi}
=\frac{\pi ^3 \varpi^3 \left(\sqrt{(\alpha +1) \left((\alpha +9) M^2-8 Q^2\right)}+3 (\alpha +1) M\right)^6}{8 (\alpha +1)^2 \left(M
   \left(\sqrt{(\alpha +1) \left((\alpha +9) M^2-8 Q^2\right)}+(\alpha +3) M\right)-2 Q^2\right)^2 \left(\exp \left(\mathcal{A}\right)-1\right)}\,,
\end{equation}
\end{widetext}
\begin{widetext}
with
\begin{align}
\mathcal{A}=\frac{\pi  \varpi 
   \left(\sqrt{(\alpha +1) \left((\alpha +9) M^2-8 Q^2\right)}+3 (\alpha +1) M\right)^3}{2 (\alpha +1) \left(M \left(\sqrt{(\alpha +1)
   \left((\alpha +9) M^2-8 Q^2\right)}+(\alpha +3) M\right)-2 Q^2\right)}.
\end{align}
\end{widetext}

 Analyzing the energy emission rate \eqref{Eq:EnergyRate} directly in terms of the parameters $\alpha$ and $Q$ reveals the following trends. Increasing the scalar-vector-tensor coupling $\alpha$ enhances the overall magnitude of the terms in the numerator, but also strengthens the exponential suppression, resulting in a slight decrease in the peak emission height while the peak frequency remains nearly unchanged. In contrast, increasing the BH charge $Q$ reduces the numerator but also decreases the exponential argument, which slightly increases the peak height while shifting the peak frequency to lower values. Therefore, the emission spectrum exhibits a delicate interplay between $\alpha$ and $Q$, with $\alpha$ primarily controlling the peak suppression and $Q$ influencing both the peak height and its position in the frequency space (see Fig. \ref{fss}).

\begin{figure}[htbp]
    \centering
\includegraphics[width=0.5\textwidth]{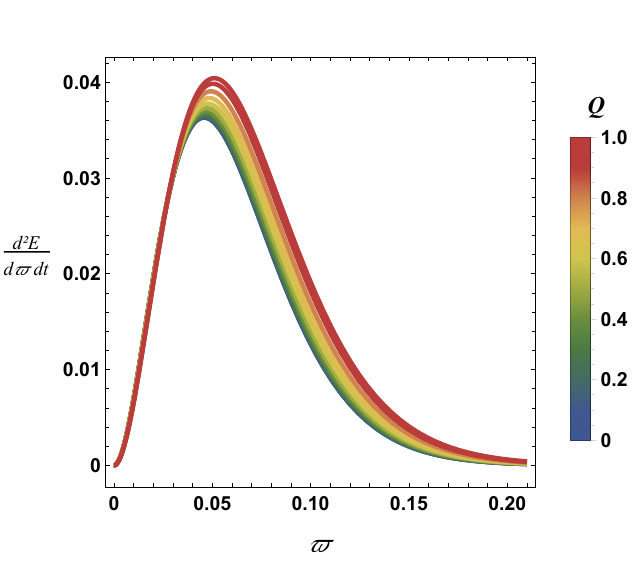}
    \vspace{0.5cm}
\includegraphics[width=0.5\textwidth]{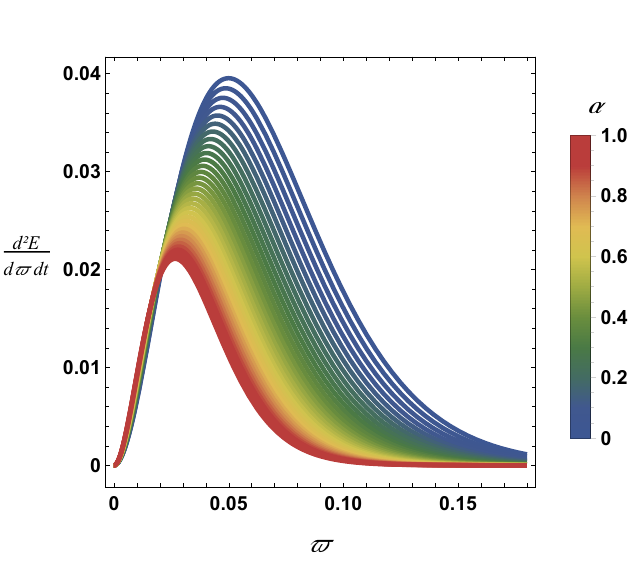}
    \vspace{0.5cm}
\caption{Emission rate of the BH  with respect to $\varpi$ using $M = 1$ for various value of the parameter space.}
    \label{fss}
\end{figure}

\section{Radiative characteristics of accretion disks}\label{isec9}
So far, null geodesics have been examined, the photon-sphere radius computed, and the corresponding BH shadow analyzed. In this section, attention is shifted to timelike geodesics, focusing on circular orbits of massive test particles. We evaluate the relevant orbital parameters, including the angular velocity, angular momentum, energy, and radius of the innermost circular orbits in the equatorial plane. These quantities play a crucial role in characterizing the radiative properties of accretion disks.

\subsection{Circular orbits}
To investigate the observational characteristics of thin accretion disks, we first analyze the motion of neutral, massive test particles forming the disk. These particles are assumed to follow stable circular orbits confined to the equatorial plane, corresponding to $\theta = \pi/2$.

For such orbits, where $\dot{r} = \dot{\theta} = \ddot{r} = 0$, the geodesic equations provide the following expression for the angular velocity of test particles in a static spacetime \cite{2017Bambi}:
\begin{equation}
\Omega=\pm\sqrt{-\frac{g_{tt,r}}{g_{\phi\phi,r}}} ,
\label{eq_omg}
\end{equation}
where $g_{ik}$ denotes the components of the metric tensor with indices $i,k = (t, r, \theta, \phi)$, while the $\pm$ sign corresponds to prograde and retrograde circular orbits, respectively. The specific energy $E/m$ and specific angular momentum $L/m$ of test particles moving along circular geodesics are expressed as:

\begin{subequations}
\begin{align}
\frac{E}{m}&=-\frac{g_{tt}}{\sqrt{-g_{tt}-g_{\phi\phi}\Omega^2}} ,
\label{12}\\
\frac{L}{m}&=\frac{g_{\phi\phi}\Omega}{\sqrt{-g_{tt}-g_{\phi\phi}\Omega^2}}.
\label{13}
\end{align}
\end{subequations}
where $m$ denotes the mass of the test particle.
These equations can be derived through various approaches, such as the Lagrangian or Hamilton–Jacobi formalisms; in all cases, the resulting expressions are equivalent.

The innermost stable circular orbit (ISCO) plays a central role in the study of particle dynamics around BHs, particularly in the framework of accretion disk physics. It corresponds to the smallest radius at which stable circular motion is possible. Once a particle moves inward past this radius, stable orbits cease to exist, and the particle inevitably plunges into the BH. Thus, the ISCO defines the inner edge of an accretion disk, beyond which the disk structure cannot be maintained. The ISCO location—and related quantities such as disk extent, temperature distribution, and luminosity—provide crucial information about the physical characteristics of BHs and their surrounding environments

The ISCO radius, $r_{\text{ISCO}}$, is obtained from the condition $dL/dr = 0$, or equivalently, $dE/dr = 0$ \cite{2016PhRvD..93b4024B}, although other equivalent formulations can also be applied to reach the same result.

For the spacetime under consideration, the orbital parameters of neutral, massive test particles take the following form

\begin{align}
\Omega &= 
\frac{\sqrt{(1+\alpha)\left[M\,(r - \alpha M) - Q^{2}\right]}}{r^{2}}, \\[1em]
\frac{E}{m} &=
\frac{r^{2} + (1+\alpha)\,(\alpha M^{2} - 2Mr + Q^{2})}
{r\,\sqrt{\,r^{2} + (1+\alpha)\,(2\alpha M^{2} - 3Mr + 2Q^{2})\,}}, \label{eq:energy}\\[1em]
\frac{L}{m} &= 
\frac{r\,\sqrt{(1+\alpha)\,\left[M(r - \alpha M) - Q^{2}\right]}}
{\sqrt{\,r^{2} + (1+\alpha)\,(2\alpha M^{2} - 3Mr + 2Q^{2})\,}},\label{eq:ang_mom}
\end{align}
\begin{equation}\label{eq:ISCO}
\frac{r_{ISCO}}{M} = \frac{B_{1}^{2} +A_{1} + (1+\alpha)(2B_{1} - 3q^{2})}{B_{1}},\qquad
\end{equation}
where
\begin{align}
B_{1} &= 
\left[B_{2} + A_{2} - A_{3} q^{2} + 2(1+\alpha)  q^{4}\right]^{1/3}, \\[1em]
B_{2} &= 
(1+\alpha)\,(\alpha+q^{2})\sqrt{(1 - q^{2})(5+\alpha - 4q^{2})}, \\[1em]
A_{1} &= 4 + 5\alpha + \alpha^{2}, \\[0.5em]
A_{2} &= 8 + 15\alpha + 8\alpha^{2} + \alpha^{3}, \\[0.5em]
A_{3} &= 9 + 14\alpha + 5\alpha^{2}\\
q&=Q/M .
\end{align}
It is worth noting that the condition $dL/dr=0$ yields three solutions. However, only the one given above represents a physically viable case.

\subsection{Accretion disk features}
To analyze the luminosity and spectral characteristics of accretion disks in the spacetime of a dilatonic dyonic BH, we employ the relativistic thin-disk model originally developed by Novikov and Thorne, and later extended by Page and Thorne for the Schwarzschild and Kerr geometries \cite{novikov1973, page1974}.

Following the methodology outlined in \cite{2021PhRvD.104h4009B,2024EPJC...84..230B,2024EPJP..139..273B,2024PDU....4601566K}, we compute the radiative flux $F$ emitted by the accretion disk, the differential luminosity $\mathcal{L}_{\infty}$ inferred from the flux, and the spectral luminosity distribution $\mathcal{L}_{\nu,\infty}$ as measured by a distant observer. The radiative flux $F$, representing the energy emitted per unit surface area per unit time, is evaluated at the ISCO and is given by:
\begin{equation}\label{eq:flux}
F=-\frac{\dot{{\rm m}}}{4\pi \sqrt{-g}} \frac{\Omega_{,r}}{\left(E-\Omega L\right)^2 }\int^r_{r_{ISCO}} \left(E-\Omega L\right) L_{,\tilde{r}}d\tilde{r},
\end{equation}
where $\dot{{\rm m}}$ denotes the mass accretion rate, which is set to unity for simplicity, and $g$ represents the determinant of the three-dimensional spatial submetric in the coordinates $(t, r, \varphi)$. In particular, the determinant is expressed as $\sqrt{-g} = \sqrt{-g_{rr} g_{tt} g_{\varphi\varphi}}$ \cite{2012ApJ...761..174B}.

In addition to the flux $F$, a related quantity of considerable astrophysical relevance is the differential luminosity $\mathcal{L}_{\infty}$, which corresponds to the energy per unit time measured by an observer at infinity \cite{novikov1973, page1974}
\begin{equation}\label{eq:difflum}
\frac{d\mathcal{L}_{\infty}}{d\ln{r}}=4\pi r \sqrt{-g}E F.
\end{equation}
In observational astrophysics, the emitted radiation is generally characterized by its frequency-dependent spectrum. Accordingly, the spectral luminosity distribution observed at infinity, denoted by $\mathcal{L}_{\nu,\infty}$, is introduced. Under the assumption of blackbody emission from the accretion disk, this distribution takes the form \cite{2020MNRAS.496.1115B}:
\begin{equation} \label{eq:speclum}
\nu \mathcal{L}_{\nu,\infty}=\frac{60}{\pi^3}\int^{\infty}_{r_{ISCO}}\frac{\sqrt{-g }E(u^t y)^4}{\exp\left[u^t y/F^{*1/4}\right]-1}dr.
\end{equation}
Here, \( u^t \) is the contravariant time component of the four-velocity, which is given by
\begin{equation}\label{eq:sample7}
u^t(r)=\frac{1}{\sqrt{-g_{tt}-\Omega^2 g_{\varphi \varphi}}},
\end{equation}
where $\Omega$ represents the angular velocity of the orbiting particles. The dimensionless parameter $y = h\nu / kT_{*}$ incorporates the Planck constant $h$, radiation frequency $\nu$, Boltzmann constant $k$, and the characteristic temperature $T_*$, which is defined through the Stefan–Boltzmann law, $F = \sigma T_*^4$, where $\sigma$ represents the Stefan–Boltzmann constant \cite{2024PDU....4601566K}.
For the spacetime under consideration, the time component of the four-velocity, $u^t$, is given by
\begin{align}
u^t &= 
\frac{r}
{\sqrt{\,r^{2} + (1+\alpha)\,(2\alpha M^{2} - 3Mr + 2Q^{2})\,}}.
\end{align}
It should be noted that, to ensure the argument of the exponential function remains dimensionless, the flux has been normalized with respect to the gravitational mass $M$, resulting in the redefinition $F^* = M^2 F$.

\subsection{Numerical results on accretion disk luminosity}

The radii of the ISCO, normalized by the BH mass $M$, are shown in the left panel of Fig.~\ref{fig:xiscoq} as a function of the charge $Q$, also normalized by $M$

\begin{figure*}[t]
\begin{minipage}{0.49\linewidth}
\center{\includegraphics[width=0.97\linewidth]{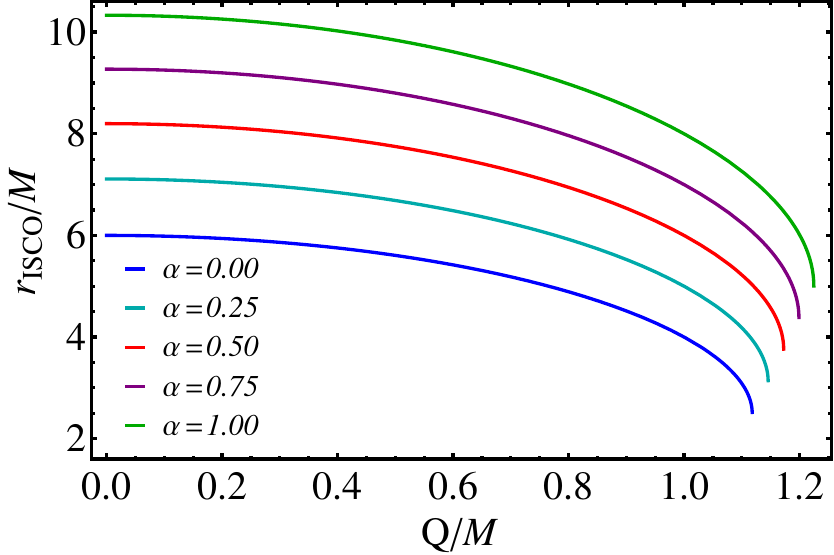}}\\ 
\end{minipage}
\hfill 
\begin{minipage}{0.50\linewidth}
\center{\includegraphics[width=0.97\linewidth]{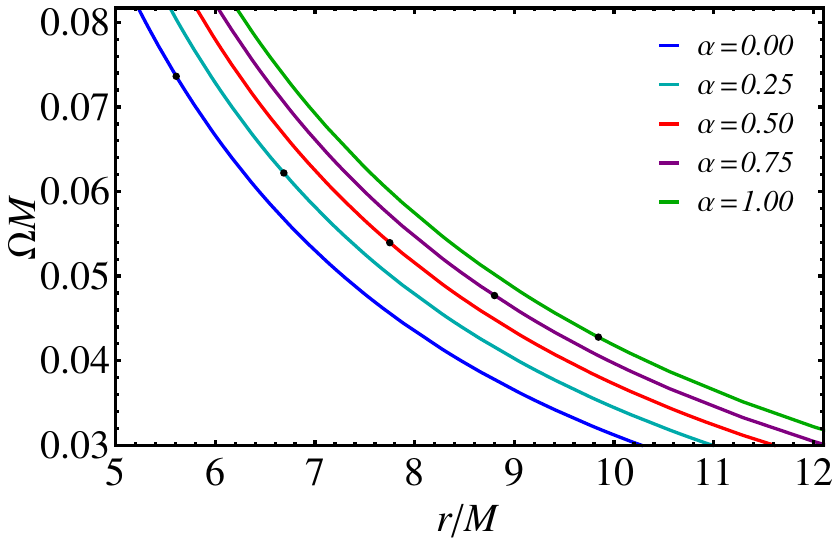}}\\ 
\end{minipage}
\caption{Left panel: The innermost stable circular orbit radii, $r_{\text{ISCO}}/M$, normalized by the gravitational mass, are presented as a function of the charge-to-mass ratio, $Q/M$.
Right panel: The dimensionless orbital angular velocity, $\Omega M$, is plotted against the normalized radial coordinate, $r/M$, for test particles in circular motion. The black dots correspond to the ISCO locations.}
\label{fig:xiscoq}
\end{figure*}

In the left panel of Fig.~\ref{fig:xiscoq}, the ISCO radius normalized by the gravitational mass, $r_{\text{ISCO}}/M$, is plotted as a function of the normalized charge, $Q/M$, for several values of the coupling parameter $\alpha = 0.0,,0.25,,0.5,,0.75,$ and $1.0$. The plot shows that $r_{\text{ISCO}}/M$ decreases with increasing $Q/M$, while it increases with larger $\alpha$, indicating that stronger charge and smaller coupling parameters correspond to more compact ISCOs.

In the right panel of Fig.~\ref{fig:xiscoq}, the dimensionless orbital angular velocity, $\Omega M$, is shown as a function of the normalized radial coordinate, $r/M$, for a fixed charge value of $Q/M = 0.5$. The black dots along the curves mark the ISCO locations. Circular orbits located to the right of each dot are stable, whereas those to the left are unstable.

\begin{figure*}[ht]
\begin{minipage}{0.49\linewidth}
\center{\includegraphics[width=0.97\linewidth]{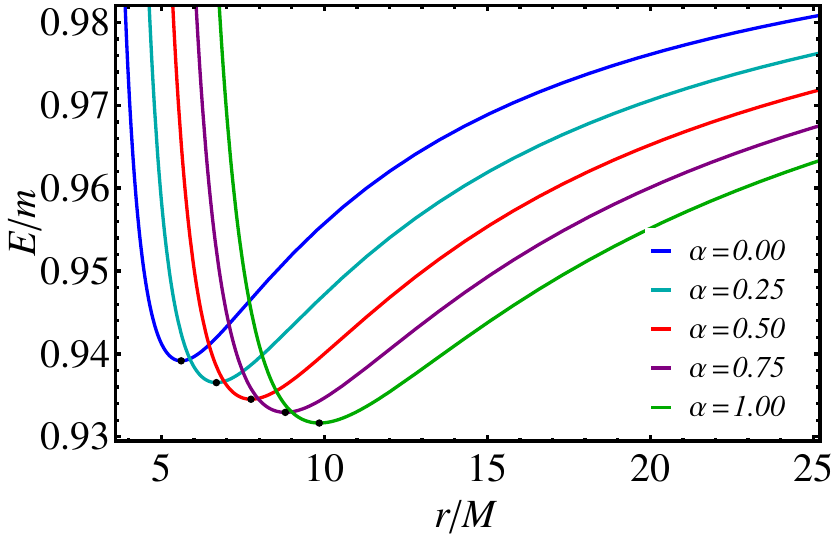}\\ }
\end{minipage}
\hfill 
\begin{minipage}{0.50\linewidth}
\center{\includegraphics[width=0.97\linewidth]{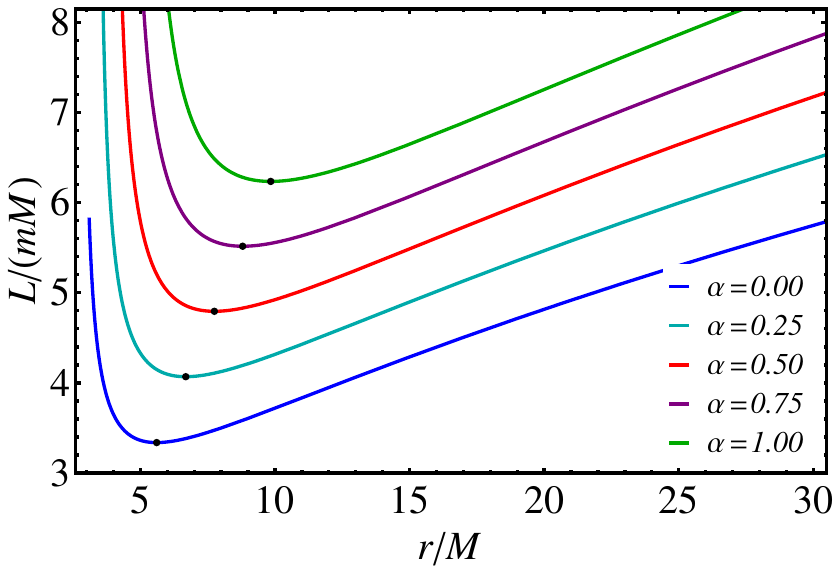}\\ }
\end{minipage}
\caption{Left panel: Specific energy, $E/m$, of test particles in circular orbits as a function of the normalized radial coordinate, $r/M$, for various values of the coupling parameter $\alpha = 0.0,,0.25,,0.5,,0.75,$ and $1.0$.
Right panel: Specific orbital angular momentum, $L/(mM)$, of test particles as a function of $r/M$ for the same set of $\alpha$ values.
In both panels, the black dots indicate the locations of the corresponding ISCO radii, $r_{\text{ISCO}}/M$.}
\label{fig:EL}
\end{figure*}
In Fig.~\ref{fig:EL}, the specific energy $E/m$ and specific angular momentum $L/(mM)$ of test particles in circular orbits are shown as functions of the normalized radial coordinate $r/M$ for a fixed value of the normalized charge, $Q/M = 0.5$. The separation between the curves corresponding to different values of the coupling parameter $\alpha$ becomes more pronounced with increasing $Q/M$.

Moreover, for all considered values of $\alpha$, the resulting solutions deviate from the Schwarzschild and Reissner–Nordstr\"{o}m geometries. Consequently, the qualitative behavior of these physical quantities remains distinctly different throughout the explored parameter space.

\begin{figure*}[ht]
\begin{minipage}{0.49\linewidth}
\center{\includegraphics[width=0.97\linewidth]{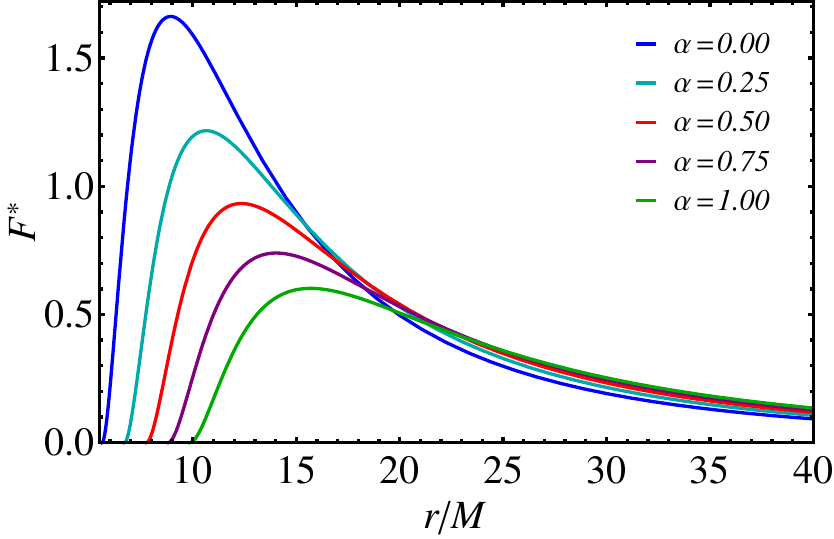}\\ } 
\end{minipage}
\hfill 
\begin{minipage}{0.50\linewidth}
\center{\includegraphics[width=0.97\linewidth]{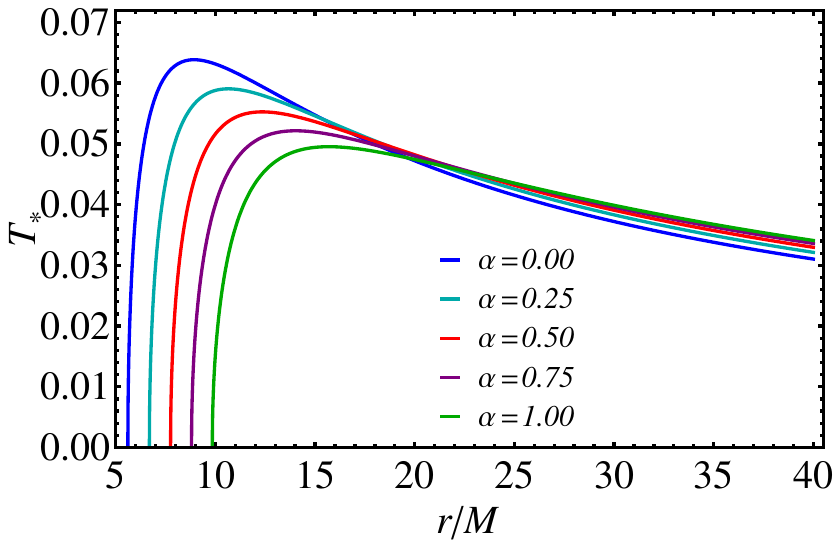}\\ }
\end{minipage}
\caption{Left panel: Rescaled radiative flux ${F}^*$ ($\times 10^5$) from the accretion disk as a function of the normalized radial coordinate $r/M$.
Right panel: Radial distribution of the radiation temperature corresponding to the emitted flux.}
\label{fig:flux_tem}
\end{figure*}

In Fig.~\ref{fig:flux_tem}, the rescaled radiative flux $F^*$ and the disk temperature $T_*$ are shown as functions of the normalized radial coordinate $r/M$, for a fixed value of the net charge $Q/M = 0.5$. Within this parametrization in terms of $M$ and $Q$, both the flux and the temperature are found to increase with the coupling parameter $\alpha$. This behavior originates from the dependence of the ISCO radius on $\alpha$ and $Q$: as $\alpha$ increases, the ISCO shifts outward, whereas larger $Q$ values cause it to contract. Consequently, the accretion disk extends closer to or farther from the BH, thereby enhancing or reducing the emitted radiation. A similar dependence is observed in the corresponding temperature distributions.

\begin{figure*}[ht]
\begin{minipage}{0.49\linewidth}
\center{\includegraphics[width=0.97\linewidth]{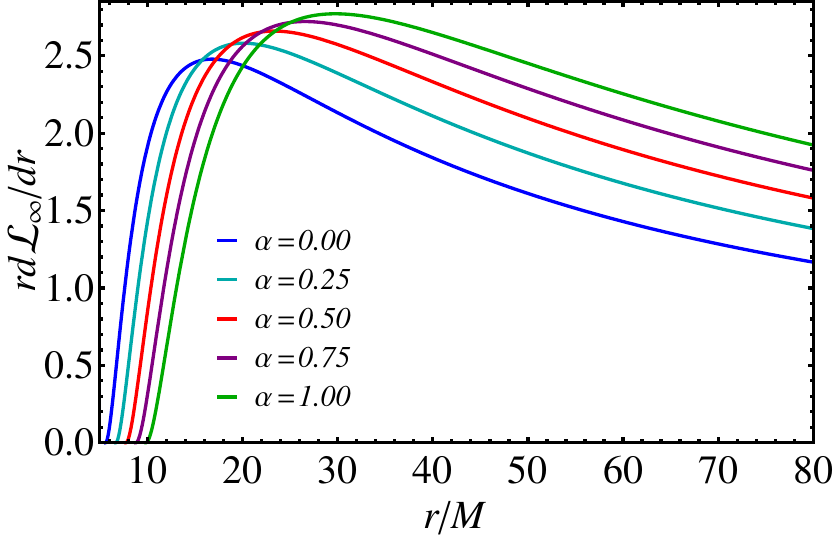}\\ } 
\end{minipage}
\hfill 
\begin{minipage}{0.50\linewidth}
\center{\includegraphics[width=0.97\linewidth]{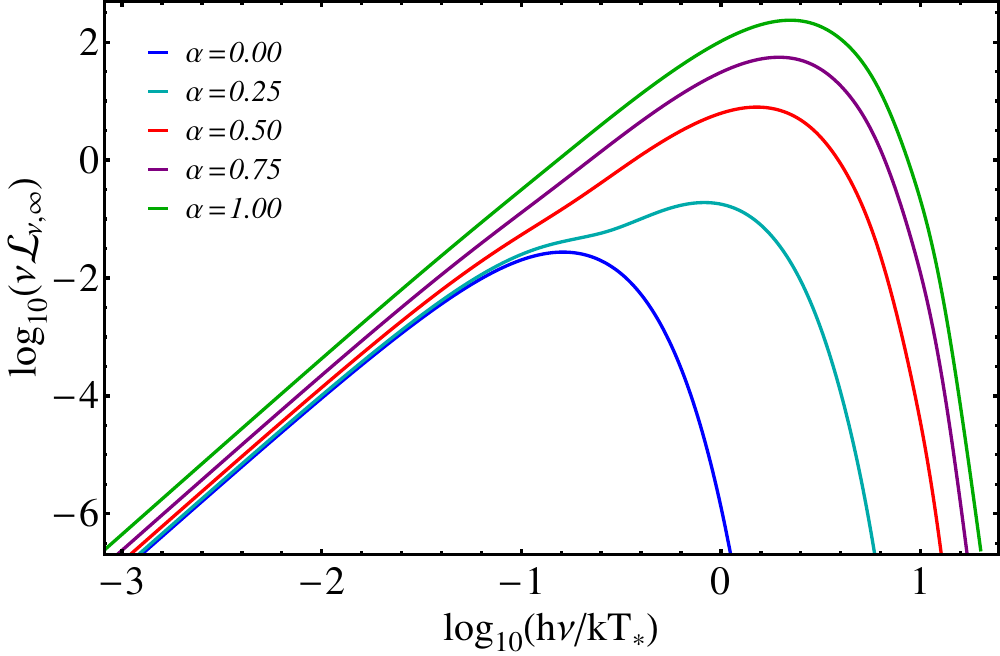}\\ }
\end{minipage}
\caption{Left panel: Rescaled differential luminosity ($\times 10^2$) versus $r/M$. Right panel: Observed spectral luminosity distribution versus $h\nu/kT_*$.
}
\label{fig:diff_spec_lum}
\end{figure*}

In Fig.~\ref{fig:diff_spec_lum}, the differential and spectral luminosities of the accretion disk are presented. The differential luminosity is plotted as a function of the normalized radial coordinate $r/M$, while the spectral luminosity is displayed as a function of the radiation frequency. Across the entire radial domain, the differential luminosity attains its maximum for $\alpha = 1$ and its minimum for $\alpha = 0$. A similar dependence is observed for the spectral luminosity, where higher values of $\alpha$ correspond to stronger emission throughout the full frequency range.

These findings emphasize an important effect: variations in the charge $Q$ and coupling parameter $\alpha$ can reproduce the radiative behavior typically attributed to Kerr BHs. Specifically, for moderate values of $\alpha$ and large $Q$, the resulting observational characteristics closely resemble those of Kerr spacetimes, effectively mimicking spin-induced signatures \cite{2021PhRvD.104h4009B,2024PDU....4601566K}. This leads to an observational degeneracy, whereby a non-rotating BH with suitable parameter combinations may appear indistinguishable from a  Kerr BH when only accretion disk emission is considered. Therefore, high-precision observational data—particularly from X-ray spectroscopy and future interferometric imaging—will be essential for resolving this degeneracy and distinguishing between rotational and scalar-tensor-vector-field–induced effects.

\subsection{Efficiency of converting matter into radiation}
The efficiency of converting accreting matter into radiation, often referred to as the binding energy per unit mass of a particle at the innermost stable circular orbit ($r_{\text{ISCO}}$), quantifies the fraction of rest-mass energy liberated before the particle plunges into the BH. By employing Eqs.~\eqref{eq:energy} and \eqref{eq:ISCO}, the efficiency can be expressed as
\begin{eqnarray}
    \eta=(1-E(r_{ISCO})/m)\times100\%
\end{eqnarray}
This quantity represents the portion of rest-mass energy released as radiation when a test particle spirals inward to $r_{\text{ISCO}}$ and subsequently crosses the event horizon. Such a high conversion efficiency plays a key role in explaining the tremendous energy output observed from compact astrophysical objects \cite{shapirobook}.

In Fig.~\ref{fig:effi}, the efficiency is plotted as a function of the normalized charge for several representative values of $\alpha$. It is observed that, for $Q/M < 1$, the efficiency increases with $\alpha$, whereas beyond $Q/M = 1$, the trend reverses, indicating a reduction in radiative efficiency for larger coupling parameters.

\begin{figure}[h]
    \centering
      \includegraphics[width=8cm]{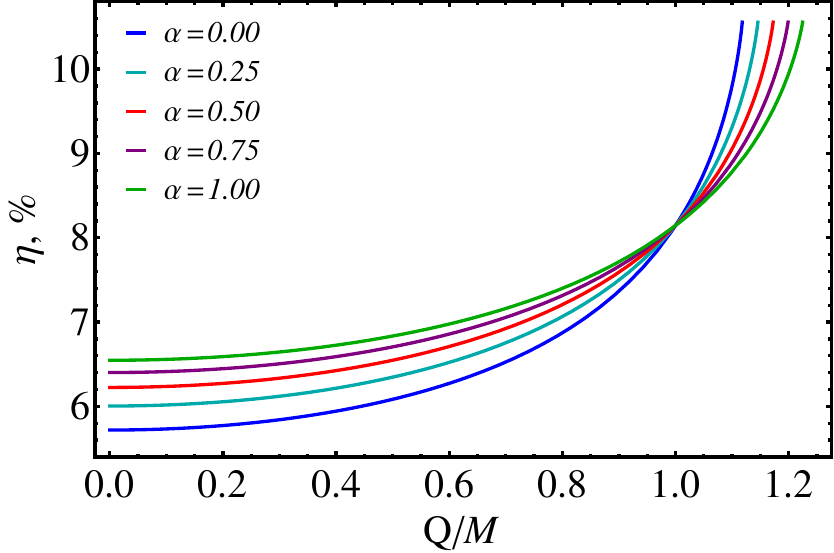}
    \caption{Efficiency versus charge}
    \label{fig:effi}
\end{figure}

\subsection{Gravitational capture cross section}
In this subsection, the gravitational capture cross section for both massive and massless particles infalling from infinity toward the BH is examined.

The capture cross section is defined as
\begin{equation}\label{eq:capture}
    \sigma_{capt}=\pi b_{max}^2
\end{equation}
where $b_{\text{max}}$ denotes the maximum impact parameter corresponding to the trajectory of a particle that becomes gravitationally captured. For non-relativistic particles, this parameter is expressed as
\begin{equation}
    b_{max}\approx L(r_{MB})/(m v_{\infty}),
\end{equation}
where $r_{\text{MB}}$ represents the radius of the marginally bound orbit, determined from the condition $E/m = 1$ in Eq.~\eqref{eq:energy}, and $v_{\infty}$ denotes the velocity of the particle at spatial infinity.

The condition $E/m = 1$ yields three mathematical solutions; however, only one of them corresponds to a physically meaningful orbit, although it formally involves a complex term
\begin{equation}
   \frac{r_{MB}}{M}= \frac{4(1+\alpha)}{3}
  + \frac{(1 - i\sqrt{3})\,D_{1}}{6\,2^{1/3}}
  + \frac{(1 + i\sqrt{3})\,D_{2}}{3\,2^{2/3}\,D_{1}} \\[6pt] \, ,
\end{equation}
where
\begin{align}
D_{1} &= \big[(1+\alpha)\,(iD_{3}-D_{4})\big]^{1/3}, \\[6pt]
D_{2} &= 4(1+\alpha)\,\big((4+\alpha)-3q^{2}\big), \\[6pt]
D_{3} &= 3(\alpha+q^{2})^{3/2}\,\sqrt{\,3(32+5\alpha)-81q^{2}\,}, \\[6pt]
D_{4} &= 128 + 112\alpha + 11\alpha^{2} - 18(8+5\alpha)q^{2} + 27 q^{4}.
\end{align}
In the limiting case where both $\alpha$ and $Q$ vanish, the expression reduces to $4M$, which corresponds to the well-known result for the Schwarzschild spacetime \cite{1972ApJ...178..347B,2025GReGr..57...91M}. In Fig.~\ref{fig:rmb}, the normalized radius $r_{\text{MB}}/M$ is plotted as a function of the BH charge for several selected values of the parameter $\alpha$. As can be seen, an increase in the BH charge tends to decrease the radius of the marginally bound orbit, whereas higher values of $\alpha$ lead to its expansion.
\begin{figure}[h]
    \centering
      \includegraphics[width=8cm]{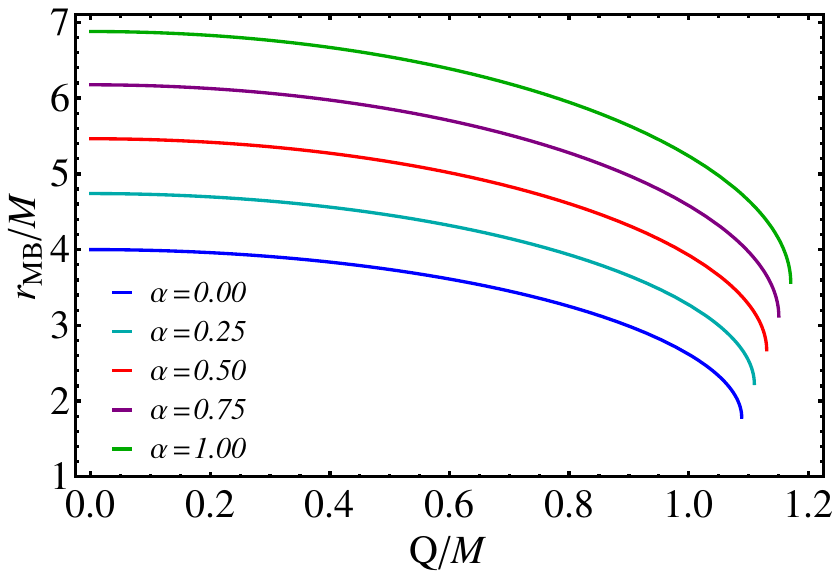}
    \caption{The radius of marginally bound orbits.}
    \label{fig:rmb}
\end{figure}

This radius is then substituted into the expression for the orbital angular momentum of the test particle (Eq.~\eqref{eq:ang_mom}). The capture cross section can be determined later in a straightforward manner. In Fig.~\ref{fig:capturetp}, the normalized capture cross section is shown as a function of the BH charge for several selected values of the parameter $\alpha$. The behavior of the capture cross section exhibits the same general trend as that of the marginally bound orbit radius, $r_{\text{MB}}$. 
%
%
%
\begin{figure}[h]
    \centering
      \includegraphics[width=8cm]{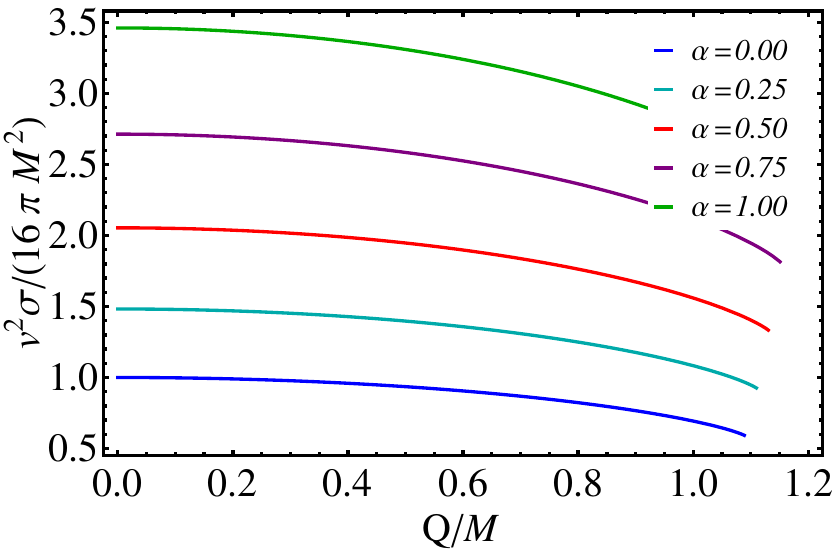}
    \caption{Gravitational capture cross section for massive test particles.}
    \label{fig:capturetp}
\end{figure}

To compute the capture cross section for massless particles, namely photons, Eq.~\eqref{eq:b_c} is employed. By setting $b_c=b_{max}$ in Eq.~\eqref{eq:capture}, the dependence of the capture cross section on the BH charge can be established for various values of the parameter $\alpha$. Figure~\ref{fig:capturephoton} presents the capture cross section normalized with respect to that of the Schwarzschild spacetime. The overall behavior of the curves in Figs.~\ref{fig:capturetp} and \ref{fig:capturephoton} is qualitatively similar. In both cases, it is evident that these physical quantities associated with test particles exhibit a strong sensitivity to the BH charge and the coupling parameter $\alpha$.

\begin{figure}[h]
    \centering
      \includegraphics[width=8cm]{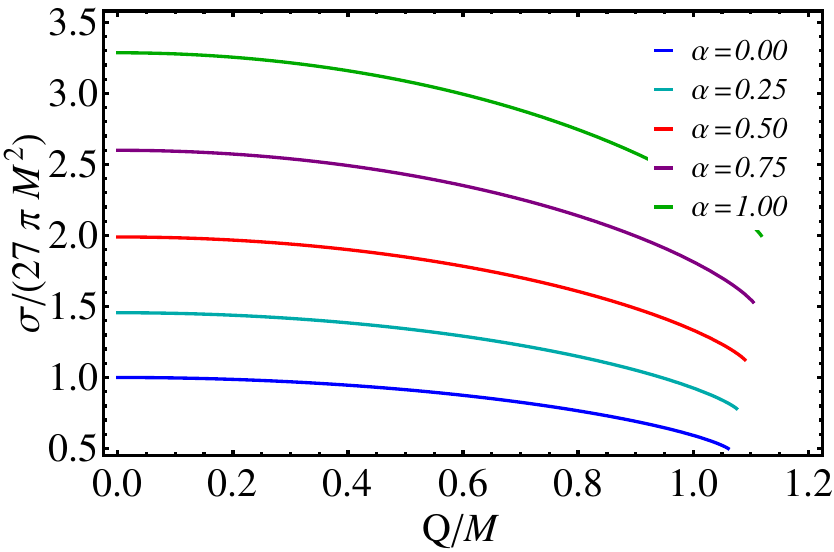}
    \caption{Gravitational capture cross section for photons.}
    \label{fig:capturephoton}
\end{figure}

\section{Discussion and Conclusion} \label{isec10}

In this work, we conducted an extensive study on charged BHs within the framework of STVG theory, analyzing their characteristics across various physical domains. Our study integrated classical and quantum perspectives to develop a unified understanding of how the STVG coupling parameter $\alpha$ and charge $Q$ affect the observable and thermodynamic characteristics of these BHs. By synthesizing results across different approaches, we established consistent patterns in how these parameters modify gravitational phenomenology in strong-field environments.

We began our analysis in Secs. \ref{isec2} and \ref{isec3} by exploring the fundamental properties of charged STVG BHs, including their horizon structure and Hawking temperature. Using the topological method, we derived the global Hawking temperature and demonstrated that the coupling parameter $\alpha$ improves the thermal stability of the system, while the electric charge $Q$ tends to suppress it. This interplay between $\alpha$ and $Q$ created a rich thermodynamic structure that persisted throughout our subsequent analyzes.

The gravitational deflection of light in Sec. \ref{isec3} was studied within the Gauss-Bonnet framework, revealing that the STVG coupling parameter $\alpha$ increases the deflection angle compared to GR. We then extended this analysis in Sec. \ref{isec4} to include plasma environments, where we found that the frequency-dependent refractive index further modifies the curvature of the optical geometry. Our results in Eq.~\eqref{eq:plasma-deflection-angle} showed that the interplay between plasma dispersion and STVG coupling creates distinctive signatures that could potentially be detected through radio-frequency lensing observations. As illustrated in Fig.~\ref{fig:beta_density}, increasing $\alpha$ progressively enhanced the deflection angle across all impact parameters and plasma densities.

In Sec. \ref{isec5}, we introduced quantum corrections through an exponential entropy model, extending the thermodynamic description beyond the semiclassical level. The corrections to internal energy (Eq.~\eqref{eq:energy_corrected}), Helmholtz free energy (Eq.~\eqref{eq:helmholtz_corrected}), pressure (Eq.~\eqref{eq:pressure_corrected}), and Gibbs potential (Eq.~\eqref{eq:gibbs_corrected}) led to modified stability conditions and shifted critical behavior in the Joule-Thomson expansion. As shown in Fig.~\ref{fig_enerji} and Fig.~\ref{fig_FF}, these corrections preserved thermodynamic consistency while introducing physically meaningful deviations at small horizon scales, indicating quantum phase transition phenomena near the Planck regime.

Our study of strong gravitational lensing in Sec. \ref{isec6} revealed that both the photon sphere radius (Eq.~\eqref{eq:rph_explicit}) and critical impact parameter (Eq.~\eqref{eq:b_c}) increase monotonically with $\alpha$ but decrease with $Q$. The logarithmic divergence in the deflection angle (Eq.~\eqref{eq:alpha_1}) near the critical impact parameter was characterized by coefficients $\bar{a}$ and $\bar{k}$ that depend explicitly on the STVG parameters, providing distinctive signatures that could be used to test this modified gravity theory. Fig.~\ref{fig:StrongDeflection} demonstrated how the deflection angle varies with impact parameter for different combinations of $\alpha$ and $Q$, showing the enhanced lensing effects with increased STVG coupling.

The BH shadow analysis in Sec. \ref{isec7} extended these results, showing that the shadow radius (Eq.~\eqref{shadow1}) grows with the coupling parameter $\alpha$, producing larger silhouettes compared to Reissner-Nordstr\"{o}m BHs. As illustrated in Fig.~\ref{fs3}, increasing $\alpha$ progressively enlarged the shadow size across all charge values, while Fig.~\ref{fs2} showed how increasing $Q$ reduced the shadow size. Using observational constraints from the Event Horizon Telescope, we found that $\alpha$ is limited to the range $0.038 < \alpha < 0.546$ for M87* and $\alpha < 0.347$ for Sgr A* across the charge range $0 < q < 1$, as shown in Fig.~\ref{fig:M87}.

In Sec. \ref{isec8}, we analyzed the energy emission spectrum, showing how the STVG parameters influence Hawking radiation. The spectral power (Eq.~\eqref{Eq:EnergyRate}) exhibited a Planck-like form with $\alpha$ primarily controlling peak suppression and $Q$ influencing both peak height and frequency position. Fig.~\ref{fss} illustrated these trends, revealing an inverse relationship between shadow size and emission intensity as a fundamental feature of BH thermodynamics in STVG theory.

From our analysis of the dynamics of the accretion disk in Sec. \ref{isec9}, we found that the radius of the innermost stable circular orbit (Eq.~\eqref{eq:ISCO}) expands with increasing $\alpha$ and contracts with larger $Q$, as shown in Fig.~\ref{fig:xiscoq}. This pattern influenced the radiative flux (Eq.~\eqref{eq:flux}), temperature distribution, and luminosity spectrum (Eq.~\eqref{eq:speclum}) of the disk. Fig.~\ref{fig:flux_tem} and Fig.~\ref{fig:diff_spec_lum} demonstrated that higher $\alpha$ values correspond to increased flux and luminosity, consistent with our findings in previous sections. Notably, we discovered that certain combinations of $\alpha$ and $Q$ can reproduce emission characteristics similar to those of Kerr BHs, introducing an observational degeneracy between vector-coupled non-rotating and rotating geometries.

We further investigated the efficiency of energy conversion in accretion processes, finding that for $Q/M < 1$, efficiency increases with $\alpha$ but beyond $Q/M = 1$, the trend reverses, as shown in Fig.~\ref{fig:effi}. Similarly, the gravitational capture cross sections for both massive particles (Fig.~\ref{fig:capturetp}) and photons (Fig.~\ref{fig:capturephoton}) exhibited consistent dependence on the STVG parameters, decreasing with charge but expanding with larger $\alpha$ values.

Throughout our analysis, we identified several consistent patterns: the STVG parameter $\alpha$  enhances gravitational effects, leading to larger photon spheres, increased deflection angles, expanded shadows, and higher accretion luminosities, while the charge $Q$ generally counteracts these effects through its repulsive electromagnetic contribution. These patterns persisted across diverse phenomena—from thermodynamics to optical properties to accretion dynamics—establishing a coherent framework for understanding charged BHs in STVG theory.

Our results provide several important pathways for testing STVG theory through astrophysical observations. The enhanced shadow sizes and modified lensing angles could be detected through high-precision imaging with next-generation interferometric arrays. The distinctive accretion disk signatures could be identified through X-ray spectroscopy of active galactic nuclei. The modified thermodynamic behavior might be probed through observations of BH mergers and their ringdown phases.

For future research, several promising directions emerge from our findings. First, extending our analysis to rotating STVG BHs would provide a more complete picture of astrophysical systems, particularly those with significant angular momentum. The Kerr-like shadows and accretion signatures we identified suggest complex interplay between rotation and STVG effects that warrants detailed investigation. Second, incorporating more realistic astrophysical environments, including magnetized plasmas, radiative processes, and non-equilibrium thermodynamics, would help bridge the gap between theoretical predictions and observational data. Finally, exploring the implications of our quantum corrections for BH information paradox and microstate counting could yield insights into fundamental physics at the interface of gravity and quantum theory.

In conclusion, our detailed analysis shows that the consistent enhancement of gravitational effects by the STVG coupling parameter $\alpha$, along with the moderating influence of charge $Q$, creates a theoretically coherent framework that can be tested through multiple independent observational channels. As astronomical capabilities continue to improve, these signatures may provide critical evidence for determining whether STVG represents a viable extension of general relativity in the strong-field regime.

}

\acknowledgments 
\.{I}.~S. and E.S. express their gratitude to T\"{U}B\.{I}TAK, ANKOS, and SCOAP3 for academic support. \.{I}.~S. also acknowledges COST Actions CA22113, CA21106, CA21136, CA23130, and CA23115 for their contributions to networking. The work of M.F. has been supported by Universidad Central de Chile through the project No. PDUCEN20240008.

\bibliography{0ref}
\bibliographystyle{apsrev}
\end{document}